\documentclass[11pt]{article}
\usepackage{amsmath}
\usepackage{amsfonts}
\usepackage{amssymb}
\usepackage{color}
\usepackage{subfigure}

\usepackage{epsfig}
\newlength{\bredde}
\def\slash#1{\settowidth{\bredde}{$#1$}\ifmmode\,\raisebox{.15ex}{/}
\hspace*{-\bredde} #1\else$\,\raisebox{.15ex}{/}\hspace*{-\bredde} #1$\fi}
\newcommand{\be}{\begin{equation}}
\newcommand{\ee}{\end{equation}}
\newcommand{\bea}{\begin{eqnarray}}
\newcommand{\eea}{\end{eqnarray}}
\newcommand{\nn}{\nonumber}
\newcommand{\svev}{\langle\langle s\rangle\rangle}

\newcommand{\la}{\lambda}
\newcommand{\ga}{\gamma}
\newcommand{\al}{\alpha}
\newcommand{\one}{\mbox{\bf 1}}
\newcommand{\bs}{\begin{split}}
\newcommand{\es}{\end{split}}

\newcommand{\sect}[1]{\setcounter{equation}{0}\section{#1}}

\def\Tr{{\mbox{Tr}}}
\def\Pf{{\mbox{Pf}}}

\begin{document}
\topmargin -1.4cm
\oddsidemargin -0.8cm
\evensidemargin -0.8cm
\title{\Large\bf Power-law deformation of Wishart-Laguerre
ensembles of\\ random matrices}

\vspace{1.5cm}
\author{~\\{\sc Gernot~Akemann} and
{\sc Pierpaolo~Vivo}
\\~\\
Department of Mathematical Sciences \& BURSt Research Centre\\
Brunel University West London,
Uxbridge UB8 3PH,
United Kingdom\\[4ex]
}

\date{}
\maketitle
\vfill
\begin{abstract}
We introduce a one-parameter deformation of the Wishart-Laguerre
or chiral ensembles of positive definite random matrices with
Dyson index $\beta=1,2$ and 4. Our generalised model has a
fat-tailed distribution while preserving the invariance under
orthogonal, unitary or symplectic transformations. The spectral
properties are derived analytically for finite matrix size
$N\times M$ for all three $\beta$, in terms of the orthogonal
polynomials of the standard Wishart-Laguerre ensembles. For large-$N$ in a
certain double scaling limit we obtain a generalised
Mar\v{c}enko-Pastur distribution on the macroscopic scale, and a
generalised Bessel-law at the hard edge which is shown to be
universal. Both macroscopic and microscopic correlations exhibit
power-law tails, where the microscopic limit depends on $\beta$
and the difference $M-N$. In the limit where our parameter
governing the power-law goes to infinity we recover the
correlations of the Wishart-Laguerre ensembles. To illustrate these
findings the generalised Mar\v{c}enko-Pastur distribution is shown
to be in very good agreement with empirical data from financial
covariance matrices.
\end{abstract}

\vfill

{\it Dedicated to the 70th birthday of Oriol Bohigas}

\thispagestyle{empty}
\newpage

\renewcommand{\thefootnote}{\arabic{footnote}}
\setcounter{footnote}{0}

%%%%%%%%%%%%%%%%%%%%%%%%%%%%%%%%%%%%%%%%%%%%%%%%%%%%%%%%%%%%%%%%%%%%%%%%%%%

%%%%%%%%%%%%%%%%%%%%%%%%%%%%%%%%%%%%%%%%%%%%%%%%%%%%%%%%%%%%%%%%%%%%%%%%%%%
\sect{Introduction}\label{intro}

Ensembles of matrices with random entries have been introduced in
the pioneering works by Wigner and Dyson in the early 1950s, in
an attempt to provide a statistical description of energy levels
of heavy nuclei based on few and minimal symmetry requirements.
Today's applications cover all areas in Physics, and we refer to
\cite{GMW} for a review. However, many years before Wigner and
Dyson, John Wishart had already introduced a random matrix
ensemble in his studies on samples from a multivariate
population. This ensemble of
random matrices can be obtained as follows: take a rectangular
matrix $\mathbf{X}$ of size $M\times N$ ($M>N$), whose entries are
independent Gaussian variables with mean zero and variance one in the real,
complex or quaternion domain, labelled by the Dyson index $\beta=1,2$ and 4,
respectively. Then form the (non-normalised)
covariance matrix of size $N\times N$,
$\mathbf{W}=\mathbf{X}^\dagger \mathbf{X}$, which is positive definite.
By comparing to such matrices 
one can establish which part of the spectrum of empirical covariance
matrices carries genuine information and which
should be discarded as contaminated by pure noise.

The ensemble of such matrices $\mathbf{W}$ is known in Random
Matrix Theory (RMT) as Wishart, Laguerre or chiral ensemble and
will be denoted by WL in the following. It has since appeared in
many different contexts, such as multivariate statistical data
analysis \cite{Johnstone}, analysis of the capacity of channels
with multiple antennae and receivers \cite{SP}, low-energy Quantum
Chromodynamics and other gauge theories \cite{SV,Jac3fold},
Quantum Gravity \cite{AKM,AMKpenner},
knowledge networks \cite{MZ1}, finance \cite{laloux} and also in
statistical physics problems, such as a class of
$(1+1)$-dimensional directed polymer problems \cite{Johansson}.
Very recent analytical results include statistics of large
deviations for the maximum eigenvalue \cite{vivo} and
distributions related to entangled random pure states
\cite{satya}.

Because of its invariance, the WL ensembles can be written in
terms of eigenvalues and completely solved, both for any finite
$N$ and $M$, as well as in the large-$N$ limit. We only give a
brief account on what is known, more details follow in the main
part later.

The WL ensemble can be exactly solved through the standard
orthogonal polynomial technique \cite{Mehta}. For a Gaussian
weight functions all spectral correlation functions can be
expressed at finite-$N$ in terms of orthogonal ($\beta=2$) or skew orthogonal
($\beta=1,4$) Laguerre polynomials, see  \cite{Mehta} for a
review. In the large-$N$ limit one has to distinguish between
smooth macroscopic, and oscillatory 
microscopic correlations on the level of mean
spacing between eigenvalues. The most known macroscopic result in
the spectral density of Mar\v{c}enko and Pastur (MP) \cite{MP},
generalising the semi-circle in the Wigner-Dyson class (WD).  On
the microscopic domain much effort has been spent on the so-called
hard edge at the origin for the positive definite eigenvalues,
which is absent in the WD class. It is governed by the Bessel-law
\cite{NagaoSlevin,SV,Forrester1st,Jacbeta1,NFbeta14} labelled by
$\beta$ and $M-N$ being finite. The robustness or universality
under polynomial deformations of the Gaussian weight function has
also been proven \cite{ADMN,Jacbeta14univ}. This perturbation
destroys the independence of the uncorrelated degrees of freedom
of $\mathbf{X}$, without changing the microscopic correlation
functions. Connected macroscopic density correlation functions
also remain universal under such deformations \cite{AKM}.

What is the motivation to further generalise WL ensembles? In the
finance and risk management domain, the empirical covariance of a
set of $N$ assets over a temporal window of size $M$ has been
under scrutiny for some time \cite{laloux}, and its eigenvalues
were shown to be distributed in reasonably good agreement with the
MP law, as if they were originated by a completely uncorrelated
data series. However, the same analysis repeated by several groups
\cite{guhr,kwapien,biroli} on different data sets have shown that
either the part of the spectrum corresponding to extremely low eigenvalues -
the most interesting for portfolio selections - or the fat tails are
not reproduced by this crude approach. This has led to
the appearance of more sophisticated models
\cite{biroli,martins,burda}, e.g. the multivariate student
distribution where the variance of each matrix entry becomes a
random variable. Another deformation of the WL ensemble has been
introduced in \cite{nagaosparse} where sparse matrices
$\mathbf{X}$ were considered. This setting has many applications
in communication theory and in complex networks (namely in the
study of spectral properties of adjacency matrices). All these
generalisations lead to a deformation of the MP law, and thus lie
outside the WL class. However, the lack of invariance in such
models generically spoils the complete solvability for all correlations
functions.

Conversely, a model with power-law tails which is exactly solvable
in principle appeared in \cite{burda}. However, the analysis in
\cite{burda} was restricted to the macroscopic spectral density,
whereas many more interesting results, and ultimately a complete
solution of the model can be obtained. This goal is achieved by
exploiting techniques and results introduced in previous works,
where the WD class was generalised using non-standard entropy
maximisation \cite{Oriol,Toscano} and super-statistical approaches
\cite{Adel,muttalib}.

We will follow these lines and provide a complete solution for all
three $\beta$ of a generalised WL model with rotational
invariance, with an emphasis on the issues of universality and
complete integrability for all spectral correlations, both
macroscopic and microscopic. After a proper rescaling and
normalisation, our $N$-independent correlations depend only on a single
parameter $\al$, given by the power-law. In the limit of
a large parameter $\al\gg1$ we recover all standard WL
correlations. For completeness, we also mention another
generalisation of WL which exploits a different direction. In
\cite{crit} the unitary WL were generalised to 
display critical statistics.

Our paper is organised as follows. In the next section
\ref{finiteN} we define our generalised WL ensembles
as a one-parameter deformation, including a general non-Gaussian
potential $V$ for all three $\beta=1,2$ and 4. The general
solution for finite-$N$ is given applying the method of (skew)
orthogonal polynomials to an integral transform of the standard
WL ensembles.

In the next section \ref{macro} we take the macroscopic large-$N$
limit for the special case of a Gaussian potential. Here macroscopic refers to
the smooth part of the spectrum, considering correlations on a distance large
compared to the mean level-spacing. We rederive a
generalisation of the semi-circle and 
MP spectral density, see refs. \cite{Oriol,Toscano,Adel} and \cite{burda}
respectively. As a new result we compute the average position of an
eigenvalue and the position of a pseudo edge. 
The former leads to the correct scaling
with $N$ and an $N$-independent
generalised spectral density displaying a power-law decay.

Section \ref{micro} is devoted to the microscopic large-$N$ limit, where
correlations on the order of the mean level-spacing are computed.
Here we can use the known universal WL results as an input to our
model. In the case when the difference of the matrix dimension
$M-N$ is kept finite the spectrum has a hard edge at the origin.
In two subsections we compute its microscopic density there, generalising
the universal Bessel-law as well as the corresponding smallest
eigenvalue distribution for all three $\beta$.

In section \ref{spacing} the nearest neighbour spacing
distribution in the bulk of the spectrum is computed using a
Wigner surmise at $N=2$ for our generalised model.

Our conclusions and outlook are presented in section \ref{con},
including a comparison of the generalised MP density to data from
financial correlation matrices. In two appendices
technical details are collected.

%%%%%%%%%%%%%%%%%%%%%%%%%%%%%%%%%%%%%%%%%%%%%%%%%%%%%%%%%%%%%%%%%%%%%%%%%%
\sect{Definition of the model and
finite-$N$ solution for general potential}\label{finiteN}

The joint probability density of our generalised Wishart-Laguerre
ensembles is defined as follows in terms of matrix elements 
\be
\label{Pga} 
\mathcal{P}_\ga[\mathbf{X}]\ d\mathbf{X}\ \propto\
\left(1+\frac{n\beta}{\gamma}\Tr
V(\mathbf{X}^\dag\mathbf{X})\right)^{-\gamma}d\mathbf{X}\ , 
\ee
where $\mathbf{X}$ is a matrix of size $M\times N$ with real,
complex or quaternion real elements for the values $\beta=1,2$ or
$4$, respectively. We define $M=N+\nu$ for later convenience,
where $\nu\geq0$ may be either finite or of order ${\cal  O}(N)$
in the large-$N$ limit. The integration measure $d\mathbf{X}$ is
defined by integrating over all independent matrix elements of
$\mathbf{X}$ with a flat measure. Expectation values of an
operator ${\cal O}$ (denoted by $\langle{\cal O} \rangle_\ga$) are
defined with respect to $\mathcal{P}_\ga$ in the usual sense.

The real positive parameter $\ga$ is to be specified below, and we
keep an additional variance-like parameter $n>0$. The so-called
potential $V$ is taken to be a polynomial of finite degree $d$,
although some of the universal results we inherit from the
WL ensembles are known to hold for a much larger class of
functions.

The measure \eqref{Pga} is well-defined and integrable only if the
following condition holds
\be
\gamma d\ >\ \frac{\beta}{2}N(N+\nu)
\label{converge}\ .
\ee
This condition, which is derived in
appendix \ref{sphere}, can be seen when changing to radial
coordinates for the matrix $\mathbf{X}^\dag\mathbf{X}$, see e.g.
\cite{ACMV,Toscano}.
In particular, the large-$N$ behaviour of any spectral property
cannot be taken for fixed $\gamma$, and a prescription about the
way both quantities should approach infinity needs to be given,
respecting the inequality eq. \eqref{converge}.

A similar model was introduced earlier generalising the
(non-chiral) Wigner-Dyson ensembles for a Gaussian potential
\cite{Oriol,Adel,Toscano}, and a similar interplay between the
deformation parameter $\gamma$ and the matrix size $N$ was
observed.

In the limit of an infinite deformation parameter
\be
\lim_{\ga\to\infty} \mathcal{P}_\ga[\mathbf{X}]\ =\ \exp[-n\beta\Tr
  V(\mathbf{X}^\dag\mathbf{X})]
\ \equiv\ \mathcal{P}[\mathbf{X}]
\label{limPga}
\ee
we recover
the standard Wishart-Laguerre or chiral ensembles
denoted by $\mathcal{P}[\mathbf{X}]$. Typically, in these ensembles one chooses
$n\sim N$ when taking the large-$N$ limit to obtain an
$N$-independent macroscopic density. In our case, this limit will
be more involved, and we will keep $n$ general for the time being.

The generalised ensembles can be related to the standard WL
through the following integral representation:
\begin{equation}
\label{GammaIdentity}
  (1+z)^{-\gamma}=\frac{1}{\Gamma(\gamma)}\int_0^\infty
  d\xi~e^{-\xi}\ \xi^{\gamma-1}\ e^{-\xi z}\ .
\end{equation}
Inserting this into the definition eq. \eqref{Pga} we obtain
\be
\mathcal{P}_\ga[\mathbf{X}]= \frac{1}{\Gamma(\gamma)}\int_0^\infty
  d\xi~e^{-\xi}\ \xi^{\gamma-1}\exp\left[-\xi\,
    \frac{n\beta}{\ga}\Tr V(\mathbf{X}^\dag\mathbf{X})\right]\ .
\label{Pgaint} 
\ee 
This relation is the key starting point to
solve our generalised model both for finite- and large-$N$. The
same trick was used for the generalisation of the Gaussian
Wigner-Dyson ensembles introduced previously in \cite{Oriol,Toscano,Adel}. In
fact, a similar technique was employed much earlier in \cite{ACMV}
when solving the fixed and restricted trace ensembles by writing
them as integral transforms of the Wigner-Dyson ensembles.

An advantage of our model over some other generalisations of WL
\cite{burda,biroli} is its invariance under orthogonal, unitary or
symplectic transformations. In particular, for $\beta=1$ 
Burda \textit{et al.} \cite{burda}
considered a very general family of probability distributions of the form: 
 $ \mathcal{P}_f[\mathbf{X}]\ \sim\  f(\Tr \mathbf{X}^T
  \mathbf{C}^{-1}\mathbf{X}\mathbf{A}^{-1})$
where $\mathbf{C}$ and $\mathbf{A}$ represent the true 
correlation and autocorrelation matrices respectively, and $f$ is a
non-negative and normalised weight function. 
Only in the special case  $\mathbf{C}=\mathbf{A}=\one$ invariance is
recovered.  
This approach has been modified in \cite{biroli} to allow for a time-dependent
random volatility.

From eq. \eqref{Pgaint} (or eq.
\eqref{Pga}) we can immediately go to an eigenvalue basis of the
positive definite matrix $\mathbf{X}^\dag\mathbf{X}$, to obtain
the following joint probability distribution function (jpdf) 
\bea
\mathcal{P}_{\ga}(\lambda_1,\ldots,\lambda_N) &\equiv&
\left(1+\frac{n\beta}{\gamma}\sum_{i=1}^NV(\lambda_i)\right)^{-\gamma}
\prod_{i=1}^N\lambda_i^{\frac12\beta(\nu+1)-1}
\prod_{j>k}^N|\lambda_j-\lambda_k|^\beta
\nn\\
&=& \frac{1}{\Gamma(\gamma)}\int_0^\infty
  d\xi~e^{-\xi}\ \xi^{\gamma-1}\ \mathcal{P}(\lambda_1,\ldots,\lambda_N;\xi)
\ .
\label{jpdEigenvalues}
\eea
It is expressed through the jpdf of the standard WL
\be\
\mathcal{P}(\lambda_1,\ldots,\lambda_N;\xi)\ \equiv\
\prod_{i=1}^N \la_i^{\frac12\beta(\nu+1)-1}
\exp\left[-\xi\,\frac{n\beta}{\ga}V(\la_i)\right]
\prod_{j>k}^N|\lambda_j-\lambda_k|^\beta
\ ,
\label{WLjpdf}
\ee
depending on $\xi$ through its weight
$\exp\left[-\xi\,\frac{n\beta}{\ga}V(\la)\right]$.
In both jpdf's
we have suppressed the constant from the integration over the angular
degrees of freedom.

For completeness we also define the corresponding partition function
\bea
\mathcal{Z}_{\ga} &\equiv&  \int_0^\infty \prod_{i=1}^N d\la_i\
\left(1+\frac{n\beta}{\gamma}\sum_{i=1}^NV(\lambda_i)\right)^{-\gamma}
\prod_{i=1}^N\lambda_i^{\frac12\beta(\nu+1)-1}
\prod_{j>k}^N|\lambda_j-\lambda_k|^\beta
\label{Zga}\\
&=&  \frac{1}{\Gamma(\gamma)}\int_0^\infty
  d\xi~e^{-\xi}\ \xi^{\gamma-1} \mathcal{Z}(\xi)\ ,
\nn 
\eea 
which is again an integral over the standard,
$\xi$-dependent WL partition function 
\be 
\mathcal{Z}(\xi)\
\equiv\ \int_0^\infty\prod_{i=1}^N d\la_i\
\la_i^{\frac{\beta}{2}(\nu+1)-1}
\exp\left[-\xi\,\frac{n\beta}{\ga}V(\la_i)\right]
\prod_{j>k}^N|\lambda_j-\lambda_k|^\beta \ . 
\label{Zxi} 
\ee
Because of this linear relation between the generalised and
standard ensembles we can immediately express all $k$-point
eigenvalue density correlation functions, denoted by $R$ for
finite-$N$, in terms of each other. They are defined in the usual
way \cite{Mehta} 
\bea 
R_\ga(\lambda_1,\ldots,\lambda_k) &\equiv&
\frac{N!}{(N-k)!} \frac{1}{\mathcal{Z}_{\ga}} \int_0^\infty
d\lambda_{k+1}\cdots d\lambda_N
\mathcal{P}_{\ga}(\lambda_1,\ldots,\lambda_N)
\label{Rkgadef}\\
&=& \int_0^\infty  d\xi~e^{-\xi}\ \xi^{\gamma-1}
\frac{\mathcal{Z}(\xi)}{\Gamma(\gamma)\mathcal{Z}_{\ga}}
R(\lambda_1,\ldots,\lambda_k;\xi)\ ,
\label{Rkgarel}
\eea
where the $k$-point correlation functions of the standard ensembles depend on
$\xi$ through the exponent in the measure
\be
R(\lambda_1,\ldots,\lambda_k;\xi)\ \equiv\
\frac{(N-k)!}{k!}  \frac{1}{\mathcal{Z}(\xi)}\int_0^\infty
  d\lambda_{k+1}\cdots d\lambda_N \mathcal{P}(\lambda_1,\ldots,\lambda_N;\xi)
\label{RkWL} \ .
\ee
The latter can be solved using the method of (skew)
orthogonal polynomials \cite{Mehta}, expressing them through the
determinant of the kernel of the orthogonal polynomials for
$\beta=2$, or the Pfaffian of the matrix kernel of skew orthogonal
polynomials for $\beta=1$ and $4$. We only detail the simpler
$\beta=2$ case here and briefly outline $\beta=1$ and $4$, where we 
refer to \cite{Mehta} for more details.

Let us define monic orthogonal polynomials and their norms for
$\beta=2$ as follows
\be
\int_0^\infty d\la
\ \la^\nu \exp\left[-\xi\,\frac{2n}{\ga}V(\la)\right] P_k(\la)P_l(\la)
\ =\ h_k \delta_{kl}\ .
\label{OPdef}
\ee
Introducing their kernel
\be
K_{N}(\la,\mu) \ \equiv\
(\la\mu)^{\frac{\nu}{2}}e^{-\xi\frac{n}{\ga}(V(\la)+V(\mu))}
\sum_{k=0}^{N-1}h_k^{-1}  P_k(\la)P_k(\mu)\ ,
\label{Kerneldef}
\ee
and applying the Christoffel-Darboux identity for $\la\neq\mu$
\be
\sum_{k=0}^{N-1}h_k^{-1}  P_k(\la)P_k(\mu)\ =\
h_{N-1}^{-1}\frac{ P_N(\la)P_{N-1}(\mu)- P_N(\mu)P_{N-1}(\la)}{\la-\mu}\ ,
\label{CD}
\ee
we can express all eigenvalue correlations of the standard ensemble
through this kernel \cite{Mehta}, 
\be
R(\lambda_1,\ldots,\lambda_k;\xi)=
\det_{1\leq i,j\leq k}\left[K_{N}(\la_i,\la_j)\right].
\ee 
We thus arrive at
\be
R_\ga(\lambda_1,\ldots,\lambda_k)
\ =\ \int_0^\infty
  d\xi~e^{-\xi}\ \xi^{\gamma-1}
\frac{\mathcal{Z}(\xi) }{\Gamma(\gamma)\mathcal{Z}_{\ga}}
\det_{1\leq i,j\leq k}\left[K_{N}(\la_i,\la_j)\right] \ ,
\label{Rkgaresult} 
\ee 
which is the main result of this section.
The simplest example is the spectral density $R_\ga(\lambda)$
given by the integral over the single kernel $K_N(\la,\la)$. Notice that it is
normalised to $N=\int_0^\infty d\la R_\ga(\lambda)$.

In order to take $N$ large we only need to know the asymptotic of
the polynomials $P_N$, take a finite determinant of size $k$ of
the asymptotic kernel and integrate once with respect to $\xi$. In
appendix \ref{beta2N} the orthogonal polynomials, the
corresponding densities and partition functions are worked out in
detail for finite-$N$ and a Gaussian potential $V(\la)=\la$ at $\beta=2$,
given
in terms of Laguerre polynomials and their norms.

As pointed out already the same result eq. \eqref{Rkgaresult}
holds for $\beta=1$ and $4$ when replacing the determinant by a
Pfaffian, $\Pf[\kappa_N(\la_i,\la_j)]$, where $\kappa_N$ is a
$2\times 2$ matrix kernel. For a Gaussian weight its skew
orthogonal polynomials are explicitly known as well in terms of
Laguerre polynomials \cite{Jacbeta1,NFbeta14}.

For completeness, we also give the partition function occurring
inside the integrand in eq. \eqref{Rkgaresult} in terms of the
norms $h_i$ of orthogonal polynomials
\be
\mathcal{Z}(\xi) \ = \
N!\prod_{i=0}^{N-1}h_i\ =\ N!\ h_0^{N}\prod_{i=0}^{N-1}r_i^{N-i}\ ,
\label{ZWLresult}
\ee
or their ratios $r_i\equiv\frac{h_{i+1}}{h_i} $. An identical result holds
for $\beta=1$ and $4$ in terms of the skew orthogonal norms \cite{Mehta}.

In the Gaussian case the ratio of partition functions
$\mathcal{Z}(\xi) /\Gamma(\gamma)\mathcal{Z}_{\ga}$ in eq.
\eqref{Rkgaresult} can be obtained most 
explicitly at finite-$N$ for all 3 values of
$\beta$ by changing to radial coordinates, see appendix
\ref{sphere} for a derivation.

%%%%%%%%%%%%%%%%%%%%%%%%%%%%%%%%%%%%%%%%%%%%%%%%%%%%%%%%%%%%%%%%%%%%%%%%%%
\sect{Macroscopic large-$N$ limit for a Gaussian potential}\label{macro}

In this section, we will restrict ourselves to the Gaussian
potential $V(\la)=\la$, deriving a generalisation of the MP
spectral density of WL for all three $\beta$. 
It exhibits a non-compact support and
power-law behaviour for large arguments, and we will distinguish
two cases. In the first subsection, we will deal with matrices
that become asymptotically quadratic with $M-N=\nu={\cal O}(1)$.
In the standard WL ensembles, at large-$N$ the spectral support is
an interval on the positive semi-axis, where the origin represents
a hard edge. 
In the second subsection, we take the limit \be
\lim_{M,N\to\infty}\frac{N}{M}\ \equiv\ c\ , \label{cdef} \ee with
$c<1$, corresponding to the case $M-N=\nu={\cal O}(N)$. In the WL
ensembles, the resulting MP macroscopic spectral density takes
support on a positive interval. Conversely, in both cases $c=1$
and $c<1$ our generalised macroscopic density will have support on
the full positive real semi-axis.

%%%%%%%%%%%%%%%%%%%%%%%%%%%%%%%%%%%%%%%%%%%%%%%%%%%%%%%%%%%%%%%%%%%%%

\subsection{Generalised semi-circle for $c=1$}

The finite-$N$ density for the WL ensembles $R(\la;\xi)$ is well
known in terms of Laguerre polynomials, see eq. \eqref{RGN} for
$\beta=2$ in the appendix \ref{beta2N}. This results into an
explicit integral representation for the spectral density of our
generalised ensembles, see eq. \eqref{RgaGN}. Despite this result,
it is rather difficult to extract
information about the macroscopic large-$N$ limit from those
analytical formulae, both for the standard and generalised WL
ensembles.

Hence, we follow here an alternative route, already exploited in
\cite{Oriol,Toscano,Adel}: we directly insert the large-$N$ result into eq.
\eqref{Rkgarel}. The $N\gg 1$ asymptotic of the WL density is
known and given by the Mar\v{c}enko-Pastur law, which is in
fact the semi-circle law in squared variables at $c=1$
\begin{equation}
\label{MPc=1N}
  \lim_{N\gg 1}R(\la)\ =\ \frac{n}{\pi}\sqrt{\frac{2N}{n\la}-1}
\ ,\ \ \mbox{with}\ \ \la\in
(0,2N/n]\ .
\end{equation}

It is given here for the Gaussian weight $\exp[-n\beta\la]$ for
all three $\beta$, and can be derived easily using the Coulomb gas
approach and saddle point method (see also \cite{NagaoSlevin}).

In order to obtain an $N$-independent macroscopic density we have
to rescale the argument of the density by the mean eigenvalue
position, $\la\to \langle \lambda\rangle x$, and divide by $N$ to
normalise the density to unity. The mean position of an eigenvalue
or first moment, $\langle \lambda\rangle_\ga$, can be computed in
the generalised model for both finite-$N$ and finite-$\nu$ (see appendix
\ref{sphere}), 
\bea 
\langle \lambda\rangle_\ga &\equiv&
\frac{\int_0^\infty d\la\ \la\ R_\ga(\la)}{\int_0^\infty d\la\
R_\ga(\la)} \ =\ \frac1N \langle\Tr
(\mathbf{X}^\dag\mathbf{X})\rangle_\ga
\nn\\
&=& \frac{\ga(N+\nu)}{2 n (\ga-\frac{\beta}{2}N(N+\nu)-1)} \ .
\label{vevla}
\eea
We will comment on the existence of this first
moment later, and we will also need this equation again when we
consider $c<1$. It correctly reproduces the known result for WL in
the limit $\lim_{\ga\to\infty}\langle \lambda\rangle_\ga= \langle
\lambda\rangle= (N+\nu)/2 n$. For the WL case we thus obtain the
following known $N$- and $\beta$-independent macroscopic density
from eq. \eqref{MPc=1N}
\begin{equation}
\label{MPc=1}
\rho(x)\ \equiv\   \lim_{N\to\infty}\frac1N\langle \lambda\rangle
R(x\langle \lambda\rangle)\ =\ \frac{1}{2\pi}\sqrt{
\frac{4}{x}-1}\ ,\ \ \mbox{with}\ \ x\in(0,4]\ .
\end{equation}
It is normalised to unity and has mean $\langle x\rangle=1$.

\begin{figure*}[htb]
\centerline{
\epsfig{file=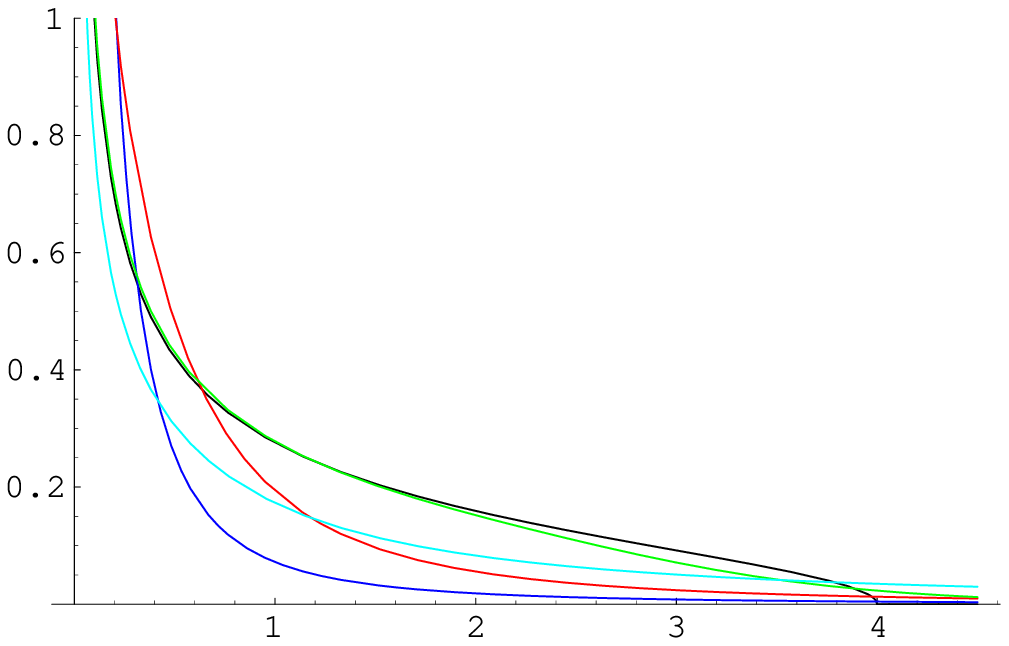,width=19pc}
  \epsfig{file=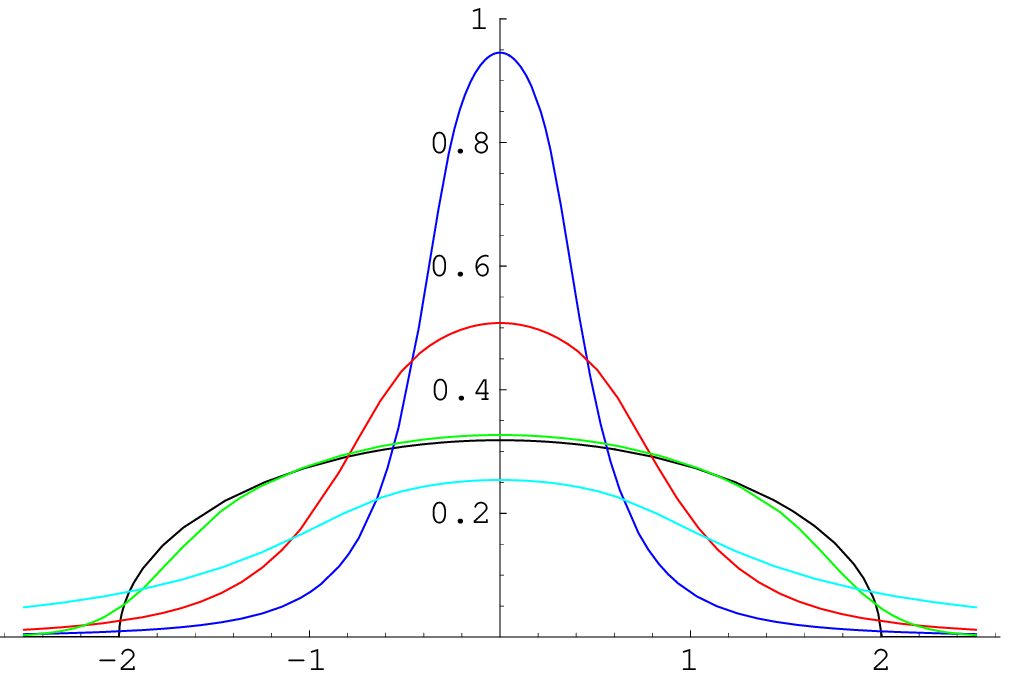,width=19pc}
\put(-100,150){$\vartheta_\al(y)$}
\put(-475,150){$\rho_\al(x)$}
\put(0,0){$y$}
\put(-250,0){$x$}
}
  \caption{
    \label{mac}
The macroscopic generalised density eq. \eqref{rhoal}
$\rho_\al(x)$ shown on the positive real line $\mathbb{R}_+$ for
$\al=-0.5,\ 0.1,\ 0.5$ and $14$ in light blue, blue, red and
green, respectively (left), and its map
$\vartheta_\al(y)=|y|\rho_\al(y^2)$ to the full real line
$\mathbb{R}$ (right). Note that the MP or semi-circle density
given in black for comparison has compact support on $(0,4]$ and
$[-2,2]$ respectively. }
\end{figure*}

We can now repeat the same steps for our generalised model, where
we will need eq. \eqref{MPc=1N} for the weight $\exp[-\xi n\beta
\la/\ga]$ (see eq. \eqref{WLjpdf}). Because of the rescaling with
respect to $\langle \lambda\rangle_\ga $ we now have to specify
the $N$-dependence of $\ga$. We keep the following combination
fixed, 
\be 
\al\ \equiv\ \lim_{N,\ga\to\infty}
\left[\ga-\frac{\beta}{2}N(N+\nu)-1\right]\ , 
\label{aldefc=1} 
\ee
with $\al>0$ finite. This satisfies the constraint
\eqref{converge}. 
We thus obtain for the generalised macroscopic density 
\bea 
\rho_\al(x) &\equiv&
\lim_{N,\,\ga\to\infty}\frac1N\langle \lambda\rangle_\ga
R_\ga(x\langle \lambda\rangle_\ga)
\nn\\
&=&  \lim_{N,\,\ga\to\infty} \frac1N\langle \lambda\rangle_\ga
\int_\mathcal{I} d\xi\ e^{-\xi}\xi^{\ga-1} \frac{{\cal
Z}(\xi)}{\Gamma(\ga){\cal Z}_\ga} \frac{n\xi}{\ga\pi}\
\sqrt{\frac{2N\ga}{n\xi x\langle \lambda\rangle_\ga}-1}\ ,
\label{rhoal1}
\eea
where the integration is restricted to the
interval $\mathcal{I}=(0,\frac{2N\ga}{nx\langle
\lambda\rangle_\ga}]=(0,4\al/x]$. In order to compute the integral
we still need the following quantity inside the integrand,
\be
\frac{{\cal Z}(\xi)}{\Gamma(\ga){\cal Z}_\ga}\ =\
\frac{\xi^{-\frac{\beta}{2}N(N+\nu)}}{\Gamma\left(\ga-\frac{\beta}{2}N(N+\nu)
\right)} \ ,
\label{Zratio}
\ee
the ratio of partition functions
given here for finite-$N$. This result is derived in appendix
\ref{sphere}, see eq. (\ref{ZratioA}).
Inserting all ingredients into eq. \eqref{rhoal1}
and taking limits with the definition \eqref{aldefc=1} we arrive
at the following result after changing variables,
\bea
\rho_\al(x)&=&
\frac{1}{2\al\pi\Gamma(\al+1)}\left(\frac{4\al}{x}\right)^{\al+2}
\int_0^1 dt\ \exp\left[-\frac{4\al}{x}t\right]\ t^{\al+1}\sqrt{\frac1t-1}\nn\\
&=&
\frac{\Gamma(\al+\frac32)}{4\al\sqrt{\pi}\Gamma(\al+1)\Gamma(\al+3)}
\left(\frac{4\al}{x}\right)^{\al+2}
\,_1F_1\left(\al+\frac32;\al+3;-\frac{4\al}{x}\right)\ .
\label{rhoal} 
\eea 
Here we have introduced the  confluent or
Kummer hypergeometric function 
\be 
_1 F_1(a;b; z)\ =\
\frac{\Gamma(b)}{\Gamma(b-a)\Gamma(a)} \int_0^1 dt\ e^{zt}\
t^{a-1}(1-t)^{b-a-1}\ . 
\label{1F1def} 
\ee 
Eq. \eqref{rhoal} is
the main result of this subsection, the macroscopic spectral
density of our generalised WL. It has an unbounded support
$(0,\infty)$, and the density as well as its first moment are
normalised to unity 
\be 
\int_0^\infty dx \rho_\al(x)\ =\ 1\ =\
\int_0^\infty dx\ x\ \rho_\al(x) \ . 
\label{norm} 
\ee 
Note that
due to this normalisation, the parameter $n$ has completely
dropped out. We are left with a one-parameter class of densities
depicted in fig. \ref{mac}, which approach the WL density for
$\al\to\infty$ as discussed below.

From the  expansion for small arguments 
$_1 F_1(a;b;z)=1+\frac{a}{b}z+\ldots$ we can immediately read off the power
law decay of our new density eq. (\ref{rhoal}) 
\be
\label{rhozetaasyp} 
\lim_{x\to\infty}\rho_\al(x)\ =\ x^{-(\al+2)}
\frac{\Gamma(\al+\frac32)}{
%2\al
\sqrt{\pi}\Gamma(\al+1)\Gamma(\al+3)}
(4\al)^{\al+1}\ (1+{\cal O}(1/x)) \ . 
\ee 
Because of $\al>0$ the
decay is always faster than quadratic. However, if we drop the
requirement for the existence of the first moment we can allow for
$\al$ to take values $-1<\al<0$ while satisfying the constraint
eq. \eqref{converge}. Keeping the same formal rescaling in eq.
\eqref{rhoal1} we arrive at the same result eq.  \eqref{rhoal} now
valid for $-1<\al$ and $\al\neq0$ \footnote{We use the obvious
notation $\al^{\al+1}=\exp[(\al+1)\ln|\al|]$.}. Our density can
thus describe power laws in between linear and quadratic decay as
well, see fig. \ref{mac}. The same feature could be incorporated
in the generalised Gaussian Wigner-Dyson model
\cite{Oriol,Adel,Toscano}.

As a check we can take the limit $\al\to\infty$ on our final
result eq. \eqref{rhoal}. This amounts to decoupling the $\gamma$-
and $N$-dependence, and thus we expect to recover the MP density
at $\gamma=\infty$.
By taking a saddle point
approximation, we find that this is indeed the case,
$\lim_{\al\to\infty}\rho_\al(x) =\rho(x)$. Hence, the density
$\rho_\al(x)$ is a well-behaved deformation of the MP density for
$c=1$.

We can also derive the behaviour of the density $\rho_\al(x)$ close
to the origin. Using the large argument asymptotic for the
confluent hypergeometric function at negative argument,
$\lim_{z\to\infty} \ _1F_1(a;b;-z)=
z^{-a}\Gamma(b)/\Gamma(b-a)(1+{\cal O}(1/z))$, we find that 
\be
\lim_{x\to0}\rho_\al(x)\ =\ x^{-\frac12} \frac{
\Gamma(\al+\frac32)}{\pi\Gamma(\al+1)\sqrt{\al}}\ (1+{\cal O}(x))\
. \label{asymp0al} 
\ee 
Therefore all our generalised densities
have a square root singularity at the origin, just as the MP
density.

To illustrate our findings we first plot eq. \eqref{rhoal} in Fig.
\ref{mac} (left) for different values of $\al$, and compare it to
the semi-circle density eq. \eqref{MPc=1}. In order to visualise the square
root singularity for all $\al$ we map the density from the
positive to the full real axis by defining 
\be
\vartheta_\al(y)\ \equiv\ |y|\rho_\al(y^2)\ , 
\label{Rmap} 
\ee
that is to a normalised density on $\mathbb{R}$,
$\int_{-\infty}^\infty dy\vartheta_\al(y)=1$. In this form it
equals the deformed semi-circle law derived from generalising the
Gaussian Wigner-Dyson ensembles in \cite{Oriol,Adel,Toscano},
where we have eliminated all irrelevant parameters.

The same map eq. \eqref{Rmap} takes the MP density eq.
\eqref{MPc=1} to the semi-circle, as mentioned already several
times,
\be
\vartheta(y)\ =\ \frac{1}{2\pi}\,\sqrt{4-y^2}\ . 
\label{semicirc}
\ee 
Both densities are shown in  Fig. \ref{mac} (right). For
$\al=14$ it already approximates the semi-circle very well,
whereas for $\al\to0$ the generalised density rapidly
approximates a $\delta$-function. For negative  $-1<\al<0$ the
height of the maximum in fig. \ref{mac} left goes down again, 
for $\al=-0.5$ the curve is even below the semi-circle.

As a final point of this subsection let us discuss the issue of
macroscopic universality of our generalised density, eq.
\eqref{rhoal}. This is a direct consequence of the
(non-)universality of the MP or semi circular density, due to the
linear relationship eq. \eqref{Rkgarel}.

The semi-circle possesses a certain weak universality in the sense
that it is the same for all three Gaussian ensembles at
$\beta=1,2,4$, as well as for independently distributed random
variables, as was shown already by Wigner. Consequently our
generalised density is universal in this weak sense, too.

On the other hand the invariant deformation of the Gaussian
potential by a polynomial in the definition of our model is
clearly {\it non-universal}, as the semi-circle becomes a
polynomial times one or several square root cuts, and we refer to
\cite{AKM} for details at $\beta=2$. Remarkably, it was found
in \cite{AKM} for $\beta=2$ that all macroscopic two- or
higher $k$-point {\it connected} density correlation functions are
universal under such perturbations $V$, depending only on a finite
number of parameters for any degree $d$. Does this universality
persist in our model? The answer is no, simply by looking at the
definition of the connected two-point density:
\be
R_\ga^{conn}(\la,\mu)\ \equiv\
R_\ga(\la,\mu)-R_\ga(\la)R_\ga(\mu)\ .
\ee
We use the same
definition for the standard WL density. It no longer relates
linearly to the corresponding connected WL two-point
density as we would subtract an integral of the product of two
1-point densities, instead of the product of two integrated
1-point densities:
\bea
R_\ga^{conn}(\la,\mu)&=& \int_0^\infty
d\xi~e^{-\xi}\ \xi^{\gamma-1}
\frac{\mathcal{Z}(\xi)}{\Gamma(\gamma)\mathcal{Z}_{\ga}}R(\la,\mu;\xi)
\nn\\
&& -
\int_0^\infty d\xi~e^{-\xi}\ \xi^{\gamma-1}
\frac{\mathcal{Z}(\xi)}{\Gamma(\gamma)\mathcal{Z}_{\ga}}R(\la;\xi)
\int_0^\infty d\xi~e^{-\xi}\ \xi^{\gamma-1}
\frac{\mathcal{Z}(\xi)}{\Gamma(\gamma)\mathcal{Z}_{\ga}}R(\mu;\xi)\nn\\
&\neq&  \int_0^\infty d\xi~e^{-\xi}\ \xi^{\gamma-1}
\frac{\mathcal{Z}(\xi)}{\Gamma(\gamma)\mathcal{Z}_{\ga}}\
R^{conn}(\la,\mu;\xi)\ .
\label{noconn}
\eea
The universal
macroscopic connected two-point function obtained in the large-$N$
limit $\rho^{conn}(x,y;\xi)$ in \cite{AKM} will thus mix with
the non-universal density $\rho(x;\xi)$, and the same feature
persists for higher $k$-point connected correlators. This
should not come as a surprise as the same situation was
encountered in the fixed or restricted trace ensembles
\cite{ACMV,AV}, being an integral transformation of the classical
Wigner-Dyson ensembles.
As observed there \cite{AV} our microscopic correlations will remain
universal, see sect.\ref{micro}.

In the next section we will study the limit $\frac{N}{M}\to c<1$.
The corresponding standard WL ensemble can be mapped to the
so-called generalised Penner model \cite{AMKpenner} with positive
definite matrices. The extra determinant from the Jacobian of this
change of variables can be written as an extra logarithm in the
potential $V\to V+N\ln|\la|$. For the same reason as given above
the universal findings made in \cite{AMKpenner} for the unitary
ensemble $\beta=2$ do not translate to the macroscopic limit in
the next section either.

%%%%%%%%%%%%%%%%%%%%%%%%%%%%%%%%%%%%%%%%%%%%%%%%%%%%%%%%%%%%%%%%%%%%%%%%%%%%

\subsection{Generalised Mar\v{c}enko-Pastur law for $c<1$}

In this subsection we deal with the limit in which the matrix $\mathbf{X}$
remains rectangular, that is both $M$ and $N=cM$ become large such
that $\lim_{N,M\to\infty} N/M=c<1$. This limit is particularly
relevant for applications to real data series.

We will follow the same steps as in the previous subsection. It is
known that in the standard WL ensembles with weight
$\exp[-n\beta\la]$ the corresponding density is given by the
following MP law,
\begin{equation}
\label{MPc<1N}
  \lim_{N\gg 1}R(\la)\ =\ \frac{n}{\pi\la}
\sqrt{\left(\la-\frac{N}{2n}X_-\right)\left(\frac{N}{2n}X_+-\la\right)}
\ ,\ \ \mbox{with}\ \ \la\in
\left[\frac{N}{2n}X_-,\frac{N}{2n}X_+\right]\ .
\end{equation}
For later convenience we have defined the bounds \be X_\pm\
\equiv\ (c^{-\frac12}\pm1)^2\ ,\ \  \mbox{with}\ \ 0<c<1\ .
\ee
In
the limit $c\to1$ we recover from eq. \eqref{MPc<1N} the
semi-circle eq. \eqref{MPc=1N} from the last section.

The $N$-independent density is again obtained after rescaling by
the mean eigenvalue position 
\be \langle \lambda\rangle_\ga \ =\
\frac{\ga N}{2 n c(\ga-\frac{\beta}{2c}N^2-1)} \ , 
\label{vevlac}
\ee 
where we have used $M=N+\nu=N/c$. For large $\ga$ we obtain
the quantity $\langle \lambda\rangle= \frac{N}{2 n c}$, which is
the average position for the standard WL in our limit $c<1$. We
thus obtain for the rescaled density MP density 
\be 
\label{MPc<1}
\rho(x)\ \equiv\   \lim_{N\to\infty}\frac1N\langle \lambda\rangle
R(x\langle \lambda\rangle)\ =\ \frac{1}{2\pi c x}
\sqrt{(x-cX_-)(cX_+-x)}\ ,\ \ \mbox{with}\ \ x\in[cX_-,cX_+]\ .
\ee 
It is normalised to unity with mean $\langle x \rangle=1$.

For the generalised model we have to insert eq. \eqref{MPc<1N},
now with weight $\exp[-\xi\frac{n\beta}{\ga}\la]$, into eq.
\eqref{Rkgarel} and rescale with respect to eq. \eqref{vevlac}. As
previously we keep fixed 
\be 
\al\ \equiv\ \lim_{N,\ga\to\infty}
\left[\ga-\frac{\beta}{2c}N^2-1\right]\ , \ \ \mbox{with}\ \al>0\ , 
\label{aldefc<1} 
\ee 
as in eq. \eqref{aldefc=1}. The rescaled
generalised density is thus given by 
\bea 
\rho_\al(x) &\equiv&
\lim_{N,\,\ga\to\infty}\frac1N\langle \lambda\rangle_\ga
R_\ga(x\langle \lambda\rangle_\ga)
\label{rhoal2}\\
&=&  \lim_{N,\,\ga\to\infty}
\frac1N\langle \lambda\rangle_\ga \int_\mathcal{I}
d\xi\ e^{-\xi}\xi^{\ga-1}
\frac{{\cal Z}(\xi)}{\Gamma(\ga){\cal Z}_\ga}
 \frac{n\xi}{\ga\pi x \langle \lambda\rangle_\ga}
\sqrt{\left(x\langle
\lambda\rangle_\ga-\frac{N\ga}{2n\xi}X_-\right)
\left(\frac{N\ga}{2n\xi}X_+-x\langle \lambda\rangle_\ga\right)} \ , \nn 
\eea 
where
$\mathcal{I}\equiv\left[\frac{c\al}{x}X_-,\frac{c\al}{x}X_+\right]$.
Filling in all definitions and changing variables we finally
arrive at the following main result of this subsection: 
\be
\rho_\al(x)\ =\ \frac{1}{2 \pi c \al \Gamma(\al+1)}
\left(\frac{c\al}{x}\right)^{\al+2} \int_{X_-}^{X_+}dt
\exp\left[-\frac{c\al}{x}t\right] t^{\al}\sqrt{(t-X_-)(X_+-t)}\ .
\label{rhoalc<1} 
\ee 
Our density in normalised to unity and has first 
moment $\langle x \rangle=1$. This result was derived previously
(modulo different notations) for $\beta=1$ in \cite{burda}, using 
different methods. The integral in eq. \eqref{rhoalc<1} can be
computed in principle in terms of a confluent hypergeometric
series in two variables (see \cite{Gradshteyn}, formulae 3.385 and
9.261(1)), but the integral form is more convenient for numerical
evaluations and an asymptotic analysis. As a first check we
recover eq. \eqref{rhoal} in the limit $c\to1$. Furthermore, one
can show that the following limit holds,
$\lim_{\al\to\infty}\rho_\al(x) =\rho(x)$, recovering the MP
density eq. \eqref{MPc<1}. Thus the large-$N$ limit and the
large-$\ga$ limit are again well behaved. We have also checked in
appendix \ref{beta2N} that the convergence with $N$ towards the
density eq. \eqref{rhoal} is very fast, see fig. \ref{mac2N}, in
fact faster than for WL. Our remark from the previous subsection allowing
$-1<\al<0$ applies here too.

\begin{figure*}[htb]
\begin{center}
  \unitlength1.0cm
\epsfig{file=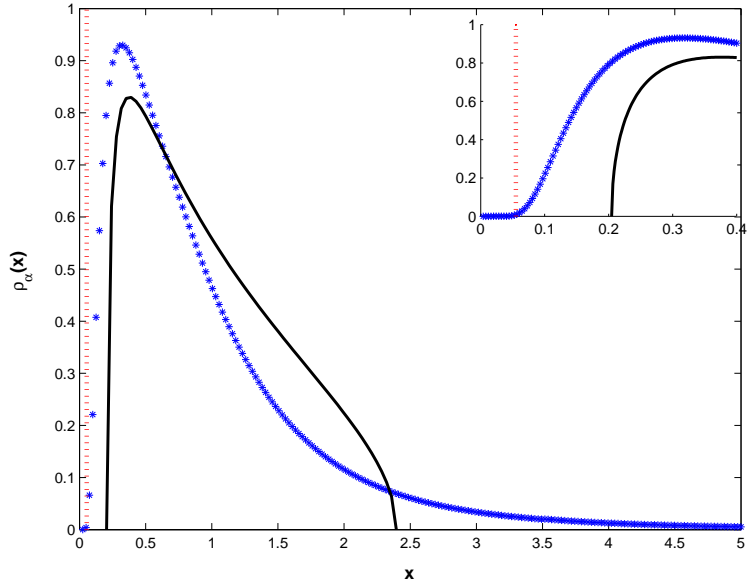,width=10cm}
  \caption{
    \label{mac2}
The macroscopic generalised density $\rho_\al(x)$  eq. \eqref{rhoalc<1}
for $\al=3$ and $c=0.3$ (blue), compared to the
MP distribution $\rho(x)$ \eqref{MPc<1} (black). In the
inset, the behaviour close to the origin is shown. The red dashed
line corresponds to the pseudo edge $\mathcal{X}_{-}$ of our
generalised MP density (see main text for details). }
\end{center}
\end{figure*}

It is easy to see how the density decays for large $x\gg1$
\be
\lim_{x\to\infty}\rho_\al(x)\ =\ x^{-(\al+2)} \frac{(c\al)^{\al+1}}{2 \pi 
%c\al
\Gamma(\al+1)}\ \mathcal{C} (1+{\cal O}(1/x))\ ,
\label{xgg1c<1}
\ee
with the same power law as for $c=1$. The
constant $\mathcal{C}$ is given by
\be
\mathcal{C}\
\equiv\ \frac18 (X_+-X_-)^2 X_-^\al \pi \
_2F_1\left(\frac32,-\al;3;-\frac{X_+-X_-}{X_-}\right)\ .
\ee
The asymptotic for small values of $x$ is less obvious to obtain, and
we find
\be
\label{asymp0}
\lim_{x\to0}\rho_\al(x)\ =\ 
x^{-\al-1/2} \exp\left[-\frac{c\al}{x}X_-\right] \mathcal{D}\ ,
\ee
where the constant $\mathcal{D}=X_{-}^{\al}
(X_{+}-X_{-})^{1/2}(c\al)^{\al-1/2}/16\Gamma(\al+1)$.

While the MP density of the standard WL ensemble eq. (\ref{MPc<1}) has compact
support between $[cX_{-}, cX_{+}]$, our generalised density is
non-vanishing on the entire real positive axis, even for $c<1$. To
the left of the edge in MP, $x<cX_-$, our density decreases but
remains non-zero. Below a certain point that we will call pseudo edge,
$\mathcal{X}_{-}$, our density becomes exponentially suppressed.
From the asymptotic \eqref{asymp0} it is possible to give an
estimate for $\mathcal{X}_{-}$, below which the density becomes
negligible. The reasoning goes as follows: writing the asymptotic
\eqref{asymp0} as  
\be 
\label{asymp0bis} \rho_\al(x)\ \sim
\exp\left[-\left((\al+1/2)\log x+\frac{c\al}{x}X_-\right)\right]
\ee
the exponential damping conventionally begins at the point
$\mathcal{X}_{-}$ where
\be
\label{damping}
(\al+1/2)\log
\mathcal{X}_{-}+\frac{c\al}{\mathcal{X}_{-}}X_-\approx 1
\ee
For the case depicted in fig. \ref{mac2} ($c=0.3$ and $\al=3$),
the above estimate reads $\mathcal{X}_{-}\approx 0.055...$, in
reasonable agreement with the inset.

%%%%%%%%%%%%%%%%%%%%%%%%%%%%%%%%%%%%%%%%%%%%%%%%%%%%%%%%%%%%%%%%%%%%%%%%%%
\sect{Universal microscopic large-$N$ limit for a general potential $V$}
\label{micro}

In this section we consider a different large-$N$ limit, the
microscopic limit, which takes us to the scale of the mean level
spacing and thus to the distribution of individual eigenvalues.
Our findings will be universal for a general non-Gaussian
polynomial potential $V(\la)$ for all three $\beta=1,2,4$,
inheriting the corresponding universality from the standard
ensembles.

We will only consider the case $c=1$ and the so-called hard edge
here, deriving a generalised Bessel-law for the microscopic
densities and their first eigenvalue distributions. For $c<1$ the
local distribution at the inner (and outer) soft edge of the standard WL
ensembles follows the Tracy--Widom-law. Although it would be very
interesting to derive the corresponding generalisation we have not
managed so far, and leave this task for future investigation.

In the first subsection the microscopic densities are derived
while the second subsection is devoted to the first eigenvalue
distributions. The matching of the two is illustrated in many
pictures throughout this section, being an important consistency check.

%%%%%%%%%%%%%%%%%%%%%%%%%%%%%%%%%%%%%%%%%%%%%%%%%%%%%%%%%%%%%%%%%%%%%%%%%%%%

\subsection{Generalised universal Bessel-law}

Let us first recall the definition of the microscopic limit in the standard WL
ensembles, resulting into the universal Bessel-law.
We will discuss in detail the case $\beta=2$ and then only quote the results
for $\beta=1,4$.

\begin{figure}[-h]
\centerline{\epsfig{figure=Crho1stal0.1b2nu0.eps
,width=13pc}
\epsfig{figure=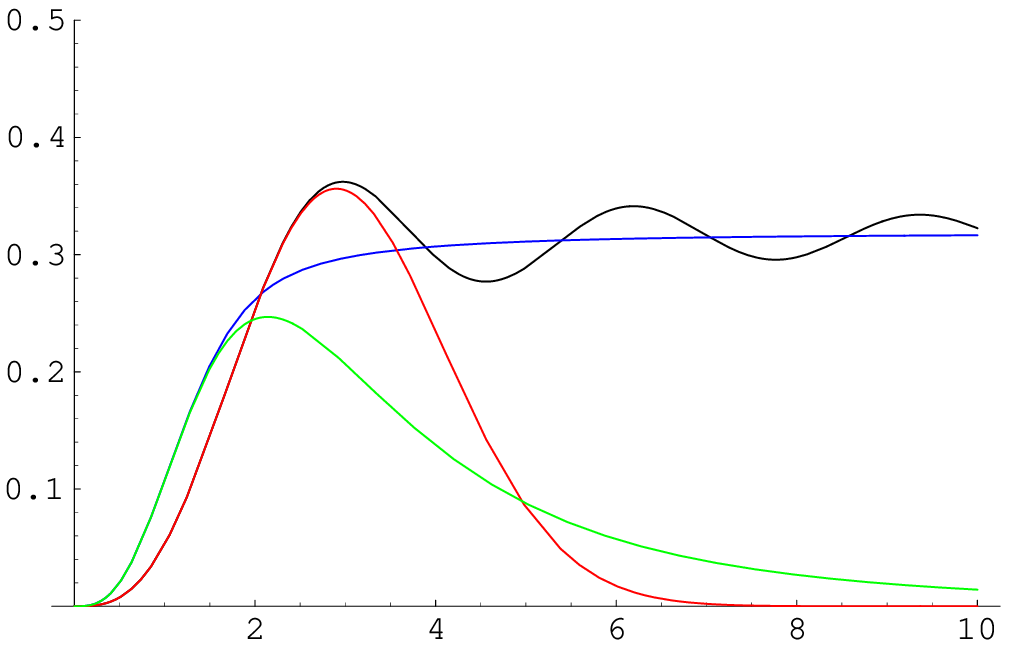
,width=13pc}
\epsfig{figure=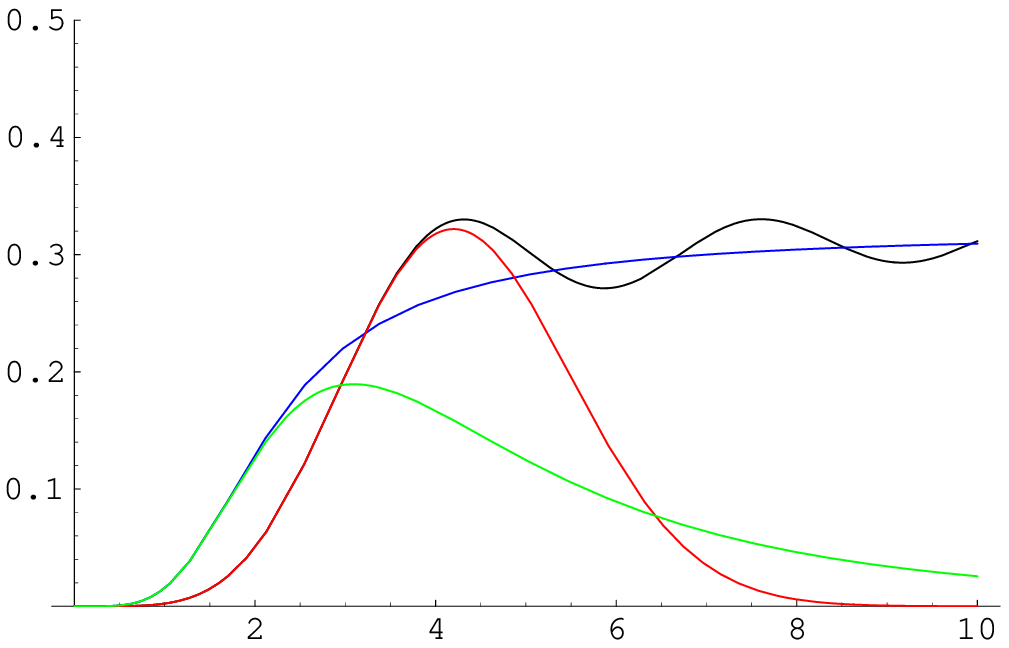
,width=13pc}
} \caption{Varying $\nu$ at $\beta=2$: the generalised microscopic
density $\vartheta_{\al,\nu}^{(2)}(y)$ eq. (\ref{genbessellaw}) 
(blue) and its first
eigenvalue (green) at
  $\al=0.1$ vs the corresponding WL Bessel density
  $\vartheta_{\nu}^{(2)}(y)$ eq. (\ref{bessellaw})
(black) and its first
  eigenvalue (red): $\nu=0$ (left), $\nu=1$ (middle) and $\nu=2$ (right).
It is clearly visible that even the first eigenvalue of the
generalised model has fat tails.} \label{varynu}
\end{figure}

For simplicity consider the Gaussian case first. Starting from the
weight $\exp[-n\beta \la]$ we first scale out the mean eigenvalue
$\la\to x\langle \la\rangle$, just as in eq. (\ref{MPc=1}) for the
macroscopic limit. On top of that we make a further rescaling by
the mean level spacing $4N^2 x= y$, keeping $y$ fixed. We
therefore define the microscopic limit as 
\be
\rho_{\nu}^{(\beta)}(y)\ \equiv\
\lim_{N\to\infty}\frac{1}{2N^2}{\langle \lambda\rangle}
R\left(\frac{y}{4N^2}\,\langle \lambda\rangle\right)\ ,
\label{micdef} 
\ee 
where $R(\la)$ is the spectral density for
finite-$N$ in one of the three WL ensembles. The result now
depends on $\beta$ and $\nu$ as indicated through the indices, in
contrast to the semi-circle law. We can then apply the following
asymptotic 
\be 
\lim_{k\to\infty}k^{-\nu}
L_k^\nu\left(\frac{z^2}{4k}\right)\ =\
\left(\frac{z}{2}\right)^{-\nu}J_\nu(z) 
\label{bessel} 
\ee 
to the
orthogonal Laguerre polynomials in the finite-$N$ density, see eq.
(\ref{RGN}) for $\beta=2$. We obtain  
\bea
\rho_{\nu}^{(2)}(y)&=& \lim_{N\to\infty} \frac{2n }{2N^2} {\langle
\lambda\rangle} \left(\frac{y\langle
\lambda\rangle2n}{4N^2}\right)^\nu\ e^{-\frac{2ny}{4N^2}\langle
\lambda\rangle} \sum_{k=0}^{N-1}\frac{k!}{(k+\nu)!}
L_k^{\nu}\left(\frac{2ny\,k}{4N^2k}\langle \lambda\rangle\right)^2
\nn\\
&=& \frac12\int_0^1dt J_\nu(\sqrt{ty})^2
\ =\  \frac12(J_\nu(\sqrt{y})^2- J_{\nu-1}(\sqrt{y})J_{\nu+1}(\sqrt{y}))\ ,
\label{bessellawsqrt}
\eea
after replacing the sum by an integral\footnote{Instead of eq. (\ref{RGN}) we
could have applied the Christoffel-Darboux identity eq.
(\ref{ChristDarb}), leading to the same result.} 
with variable $t=k/N$. The Bessel
density is plotted for different values of $\nu$ in fig. \ref{varynu} together
with the first eigenvalue from the next subsection,
after changing to the conventional squared
variables $y\to y^2$ (see eq. (\ref{Rmap})),
\be
\vartheta_{\nu}^{(2)}(y)
\ =\ \frac{|y|}{2}(J_\nu(y)^2 - J_{\nu-1}(y)J_{\nu+1}(y))\ .
\label{bessellaw}
\ee
This result is universal \cite{ADMN}
being valid for any potential $V$ with spectral
support including the origin. We only
have to rescale in eq. (\ref{micdef})
by the macroscopic density in terms of the squared variables
$\pi\vartheta(0)$ for a general potential $V$,
instead of the Gaussian macroscopic density eq. (\ref{semicirc}) where
$\pi\vartheta(0)=1$.
In other words all orthogonal polynomials eq. (\ref{OPdef}) tend to Bessel-$J$
functions in the microscopic limit.

We can now repeat the above analysis for our generalised
microscopic density. The scaling of $\ga$ with $N$ keeping
$\al$ fixed is kept throughout this entire section. In the
previous section we found that the density diverges exactly like
the standard density as an inverse square root, see eq.
(\ref{asymp0al}). For that reason the microscopic rescaling 
is the same, without changing powers of $N$.
However, the constant in front of the {\it macroscopic} density at
the origin is not the standard Gaussian result,
$1/\pi=\vartheta(0)$, but given by eq. (\ref{asymp0al}),
$\vartheta_{\al}(0)=b/\pi$ with 
\be 
b\ \equiv\ \frac{
\Gamma(\al+\frac32)}{\Gamma(\al+1)\sqrt{\al}}\ . \ee We therefore
define the microscopic limit as \be \rho_{\al,\,\nu}^{(\beta)}(y)
\ \equiv\   \lim_{N,\ga\to\infty}\frac{1}{2N^2b^2} {\langle
\lambda\rangle_\ga} R_\ga\left(\frac{y}{4N^2b^2}\,\langle
\lambda\rangle_\ga\right)\ . 
\label{micaldef} 
\ee

Because of the universality we just stated we can restrict ourselves to
do the computation for the orthogonal polynomials with
Gaussian weight $\exp[-2n\xi\la/\ga]$ in eq. (\ref{OPdef}).
Taking the microscopic limit eq. (\ref{micaldef}) and
inserting eq. (\ref{bessellawsqrt}) we obtain
\be 
\vartheta_{\al,\,\nu}^{(2)}(y)= \frac{1}{\Gamma(\al+1)}
\int_0^\infty  d\xi~e^{-\xi}\ \xi^{\al}  \frac{\xi|y|}{2{\al}b^2}
\left(J_\nu\Big(\frac{y}{b}\sqrt{{\xi }/{\al}}\Big)^2-
J_{\nu-1}\Big(\frac{y}{b}\sqrt{{\xi }/{\al }}\Big)
J_{\nu+1}\Big(\frac{y}{b}\sqrt{{\xi }/{\al }}\Big) \right).
\label{genbessellaw} 
\ee 
This is the first main result of this
subsection given here in terms of squared values. The integral could be
expressed in terms of generalised hypergeometric functions, but for plots this
representation is preferable.
Note that in our the calculation we have inserted the ratio
of partition functions eq. (\ref{Zratio}) for the Gaussian models.
This quantity is again universal as in the large-$N$ limit not
only the polynomials themselves, but also their norms become
universal. Starting from eq. (\ref{ZWLresult}) the standard WL
partition functions is given as follows \be
\lim_{N\to\infty}(\log[\mathcal{Z}(\xi)] - \log[N!\,h_0^N])\ =\
\lim_{N\to\infty}N\sum_{i=0}^{N-1}\left(1-\frac{i}{N}\right)
\log[r_i] \ =\ \int_0^1dt (1-t)\log[r(t)] \label{logZWL} \ee The
$\xi$-dependent ratios of the norms determined by the so-called
string or recursion equation at finite-$N$ have a universal limit
$r(t)$ \cite{ADMN}. Re-exponentiating and inserting this universal
result eq. (\ref{logZWL}) into eq. (\ref{Zga}), the generalised
partition function ${\cal Z}_\ga$ also becomes universal in the
large-$N$ limit, and thus the ratio eq. (\ref{Zratio}) as well.
The constant factor that we have subtracted on the left hand side
of eq. (\ref{logZWL}) cancels out when taking the ratio.

\begin{figure}[-h]
\centerline{\epsfig{figure=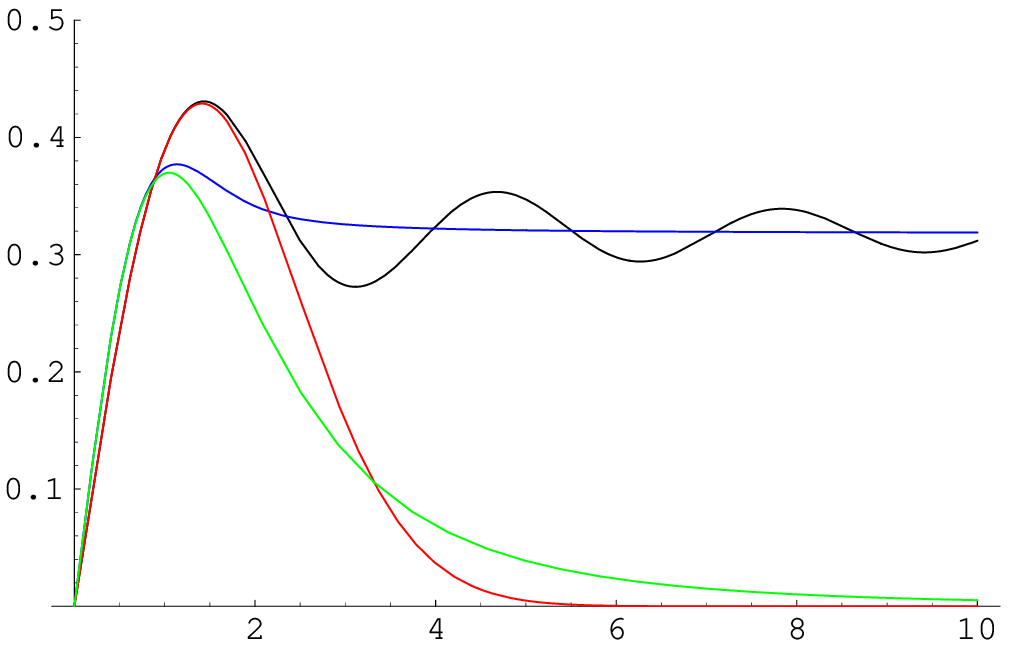,width=13pc}
\epsfig{figure=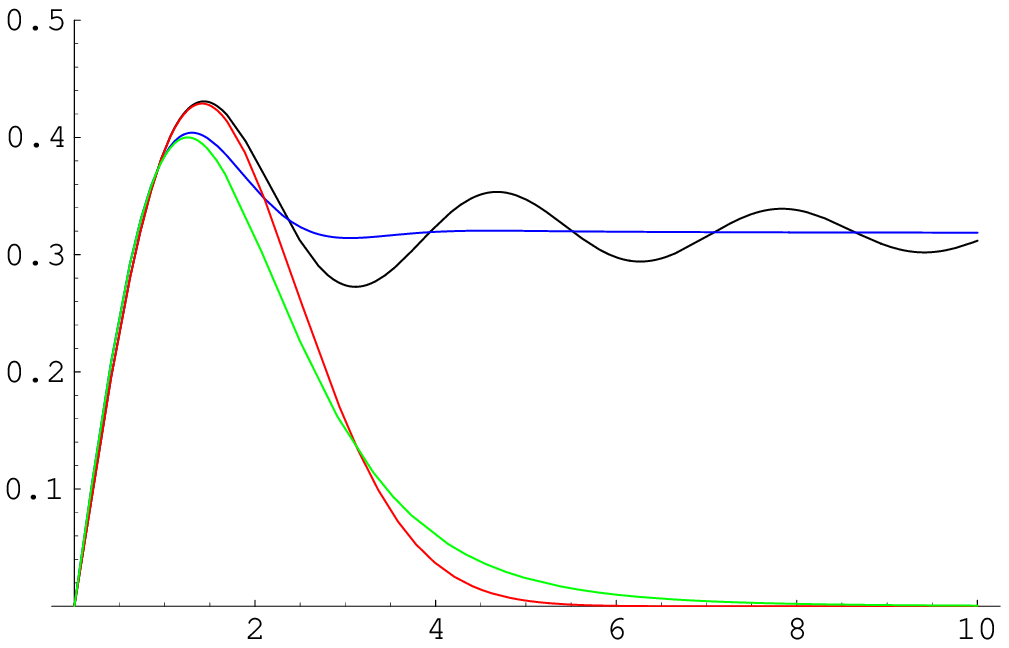,width=13pc}
\epsfig{figure=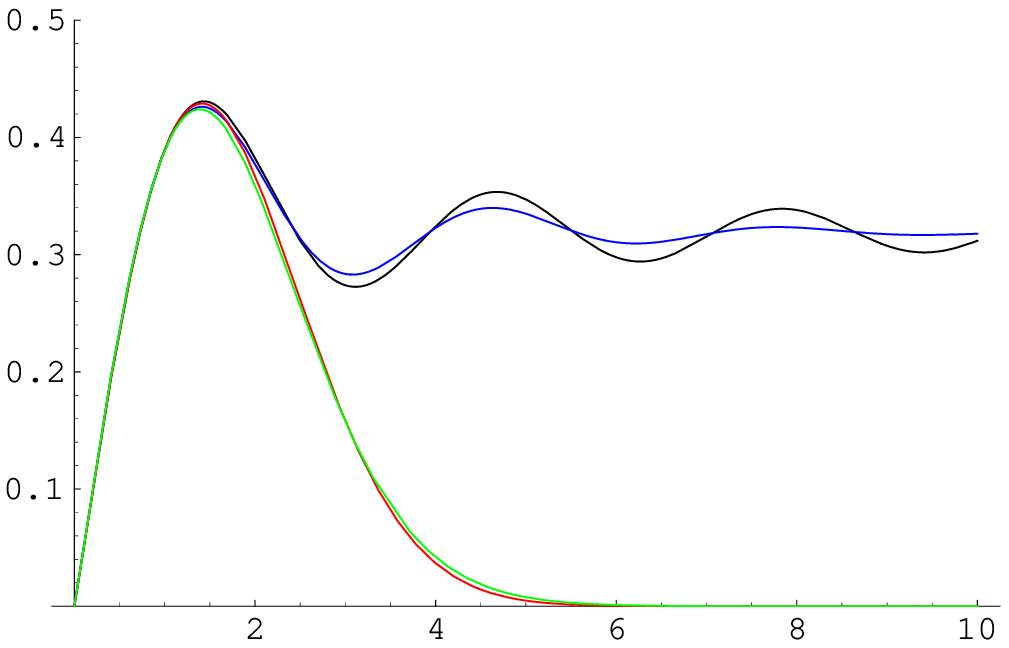,width=13pc}
} \caption{Varying $\al$ at   $\beta=2$: the generalised
microscopic density $\vartheta_{\al,\nu}^{(2)}(y)$ eq. (\ref{genbessellaw})
(blue) and its
first eigenvalue (green) at
  $\al=0.1$ (left),  $\al=2$ (middle),  and
  $\al=20$ (right),
vs the corresponding WL Bessel density
  $\vartheta_{\nu}^{(2)}(y)$ eq. (\ref{bessellaw})
(black) and its first
  eigenvalue (red).}
\label{varyal}
\end{figure}

The new microscopic density eq. (\ref{genbessellaw}) generalising
the Bessel-law eq. \eqref{bessellaw} 
is shown in figs. \ref{varyal} and \ref{varynu} for
various values of $\al$ and $\nu$. 
The plots include the corresponding first eigenvalues to be derived
later.
As a check we can take the
limit $\al\to\infty$ to analytically reobtain eq.
(\ref{bessellaw}) from eq. (\ref{genbessellaw}). This is
illustrated in fig. \ref{varyal} where we observe that the
convergence is rather slow. We have checked that for $\al\sim
{\cal O}(150)$ the first three maxima become indistinguishable.

The procedure for $\beta=1$ and $4$ is the same as above and so we
can be more concise. The corresponding microscopic densities of
the WL ensembles are again universal \cite{Jacbeta14univ}.
Computed initially in \cite{Jacbeta1,NFbeta14} the obtained
expressions can be simplified. They can be expressed through the
$\beta=2$ density eq. (\ref{bessellaw}) plus extra terms as shown for
$\beta=1$ \cite{FNHbeta1} and $\beta=4$ \cite{BBJMVWbeta4},
\bea
\vartheta_{\nu}^{(1)}(y)&=& \vartheta_{\nu}^{(2)}(y) +\frac{1}{2}
J_\nu({|y|})
\left( 1-\int_0^{{|y|}} dt J_\nu(t) \right),
\label{rhomic1}\\
\vartheta_{\nu}^{(4)}(y) &=& \vartheta_{2\nu}^{(2)}(2y)
-\frac{1}{2}J_{2\nu}(2{y}) \int_0^{2{|y|}} dt J_{2\nu}(t)\ .
\label{rhomic4} 
\eea 
The generalised densities immediately follow.
Because of the linear relationship they are also expressed through
the generalised $\beta=2$ density eq. (\ref{genbessellaw}):
\bea
\vartheta_{\al,\,\nu}^{(1)}(y)&=& \vartheta_{\al,\,\nu}^{(2)}(y)
+ \frac{1}{\Gamma(\al+1)}
\int_0^\infty\!\!  d\xi\ e^{-\xi}\ \xi^{\al}
 \sqrt{\frac{\xi}{\al b^2}}\frac12
J_\nu\Big(\frac{|y|}{b}\sqrt{\xi /\al}\Big)
\Big( 1-\int_0^{|y|\sqrt{\xi /\al b^2}}\!\!\!\!\!\! dt J_\nu(t) \Big)\!,
\label{rhomical1}\\
\vartheta_{\al,\,\nu}^{(4)}(y)
&=& \vartheta_{\al,\,2\nu}^{(2)}(2y) -
\frac{1}{\Gamma(\al+1)}
\int_0^\infty\!\!  d\xi\ e^{-\xi}\ \xi^{\al}
 \sqrt{\frac{\xi}{\al b^2}}\frac12
J_{2\nu}\Big(\frac{2y}{b}\sqrt{\xi /\al}\Big)
\int_0^{2|y|\sqrt{\xi /\al b^2}}\!\!\! dt J_{2\nu}(t).
\label{rhomical4}
\eea
For a given $\nu$ the inner
integral over the single Bessel-$J$ function can be performed analytically. It
is given terms of Bessel-$J$
functions for odd values of $\nu$, e.g. $\int_0^{v}
dt J_1(t)=J_0(v)$, and additional Struve functions for even $\nu$.

The generalised microscopic densities
$\vartheta_{\al,\,\nu}^{(\beta)}(y)$ are compared to the standard
ones below in fig. \ref{varynubeta1} for $\beta=1$, and in fig.
\ref{varynubeta4} for $\beta=4$.

Higher order correlation functions can be computed along the same lines by
inserting the asymptotic Bessel kernels into eq. (\ref{Rkgaresult}), and we
only quote the simplest final result for $\beta=2$:
\bea
\vartheta_{\al,\nu}^{(2)}(y_1,\ldots,y_k)
&=& \frac{\prod_{i=1}^k\frac12 |y_i|}{\Gamma(\al+1)}\int_0^\infty
  d\xi~e^{-\xi}\ \xi^{\al}  \prod_{j=1}^k\sqrt{\frac{\xi}{\al}}\frac{|y_j|}{b}
\nn\\
&&\times\det_{1\leq i,j\leq k}\left[
\frac{
J_\nu\Big(\frac{y_i}{b}\sqrt{{\xi }/{\al}}\Big)
J_{\nu+1}\Big(\frac{y_j}{b}\sqrt{{\xi }/{\al}}\Big)
-(i\leftrightarrow j)
}{y_i-y_j}
\right] \ .
\label{rhoalkmic}
\eea
The corresponding results for $\beta=1,4$ are given in terms of a Pfaffian of
a matrix kernel \cite{NFbeta14}, and for a
discussion of a relation between the three universal kernels we refer to
\cite{Jacbeta14univ}.

A feature we observe for all three $\beta$  is that for $\al\leq
{\cal O}(1)$ the oscillations of the Bessel density are completely
smoothed out, apart from the first peak. A similar feature was
observed in a generalisation of the unitary WL ensemble for
critical statistics \cite{crit}. However, no power law tails seem
to be present in such a model where only the generalised
microscopic density and number variance were computed.

It is known that for standard WL the maxima of the Bessel density
correspond to the location of individual eigenvalues \cite{DN}, as we will
see in the next subsection.
On the other hand the microscopic density of the WL ensembles in
the bulk is completely flat, equalling $\frac1\pi$ in our
normalisation. We may thus suspect that in the generalised model
the bulk is approached much faster than in the standard WL, where
localised maxima persist to $y\gg10$.
We therefore focus mainly on the first eigenvalue distribution in
the generalised model which is the subject of the next subsection.

%%%%%%%%%%%%%%%%%%%%%%%%%%%%%%%%%%%%%%%%%%%%%%%%%%%%%%%%%%%%%%%%%%%%%%%%%%

\subsection{Generalised universal
first eigenvalue distribution at the hard edge}

The probability that the interval $(0,s]$ is empty of eigenvalues is
defined as follows,
\bea
E_{\ga}(s) &\equiv&
\frac{1}{\mathcal{Z}_{\ga}} \int_s^\infty d\lambda_{1}\cdots d\lambda_N
\mathcal{P}_{\ga}(\lambda_1,\ldots,\lambda_N)
\label{Egadef}\\
&=& \int_0^\infty  d\xi~e^{-\xi}\ \xi^{\gamma-1}
\frac{\mathcal{Z}(\xi)}{\Gamma(\gamma)\mathcal{Z}_{\ga}}
E(s;\xi)\ ,
\label{Egarel}
\eea
where the gap probability of the WL ensembles is defined as
\be
E(s;\xi)\ \equiv\
\frac{1}{{\cal Z}(\xi)}
\int_s^\infty
  d\lambda_{1}\cdots d\lambda_N \mathcal{P}(\lambda_1,\ldots,\lambda_N;\xi)
\label{EWL} \ . 
\ee 
Both quantities are normalised to unity at
$s=0$ and vanish at $s=\infty$. The distribution of the first
eigenvalue $p(s)$ simply follows by differentiation. 
\be
p_{\ga}(s)\ \equiv\ -\frac{\partial}{\partial s}{E}_\ga(s)\ ,
\label{p1def} 
\ee 
and likewise for WL. In WL the gap probability
$E(s)$ and the first eigenvalue distribution $p(s)$ are explicitly
known and universal in the microscopic large-$N$ limit for all
$\nu$ at $\beta=2$, for odd values of $\nu$ and 0 at $\beta=1$, and for
$\nu=0$ at $\beta=4$. This has been shown by various authors
independently \cite{Forrester1st,WGW,DNW,DN}. In some cases only
finite-$N$ results are know in terms of a hypergeometric function
of a matrix valued argument \cite{dumitriu,edelman}, from which
limits are difficult to extract.

Although $p(s)$ follows from $E(s)$,
the most compact universal formulas are known
directly for $p(s)$ for all three $\beta$ \cite{DN}.
There, also the second and higher eigenvalue
distributions are given, which we will not consider here.

For pedagogical reasons we start once more with an explicit
calculation for $\beta=2$ and the Gaussian ensemble. At $\nu=0$
the pre-exponential factor is absent and we have for WL with
weight $\exp[-2n\la]$  
\be 
E(s)\ =\ \frac{1}{{\cal Z}}
\int_s^\infty
  d\lambda_{1}\cdots d\lambda_N
\exp\left[-2n\sum_{i=1}^N\la_i\right]
\prod_{j>k}^N|\lambda_j-\lambda_k|^2 \ =\ \exp[-2nNs]\ ,
\label{Eb2nu0} 
\ee 
shifting all integration variables by $s$ and
using the invariance of the Vandermonde determinant. This result
is exact for any $N$ and identical to the properly rescaled
large-$N$ result when keeping $Ns$ fixed.

The generalised ensemble follows easily, by inserting this result
into eq. (\ref{Egarel}) 
\be 
E_\ga(s)\ =\ \int_0^\infty
d\xi~e^{-\xi}\ \xi^{\gamma-1}
\frac{\mathcal{Z}(\xi)}{\Gamma(\gamma)\mathcal{Z}_{\ga}}
\exp\left[-\frac{2n\xi Ns}{\ga}\right] \ =\ \left(
1+\frac{2nNs}{\ga}\right)^{-(\ga-N^2)}\ , 
\ee 
which is also exact
for finite and infinite $N$. This very fact implies that we have
full control of the large-$N$ limit. Because of the fat tails in
our distributions it was not a priori clear if the scales at the
hard edge and in the bulk would mix. This implicit assumption of a
separation of scales in the previous subsection is thus fully
justified.

The microscopic limit can be taken following eq. (\ref{micdef})
for WL: 
\be 
{\cal E}^{(\beta)}_\nu(y)\ \equiv\
\lim_{N\to\infty}
E\left(\frac{y}{4N^2}\langle\la\rangle\right)\ , 
\ee 
where we
explicitly indicate the dependence on $\beta$ and $\nu$, as in the
previous subsection. As a result we obtain for $\beta=2$ and
$\nu=0$ 
\be 
{\cal E}^{(2)}_{\nu=0}(y)\ =\ \exp\left[-\frac14 y\right]\ , 
\label{Emic} 
\ee 
and for the corresponding generalised
gap probability 
\be 
{\cal E}^{(2)}_{\al,\,\nu=0}(y)\equiv\
\lim_{N,\ga\to\infty}
E_\ga\left(\frac{y}{4N^2b^2}\langle\la\rangle_\ga\right) \ =\
\left(1+\frac{y}{4 \al b^2}\right)^{-(\al+1)} \ .
\ee 
The first
eigenvalue distribution can be compared to the microscopic
densities $\vartheta_\al(y)$ in squared variables:
\be 
\wp_{\nu=0}^{(2)}(y)\ \equiv\ -\frac{\partial}{\partial
y}{\cal E}^{(2)}_0(y^2) \ =\ \frac12 |y| \exp\left[-\frac14
y^2\right] \ , 
\label{wp1def} 
\ee 
for WL, and for the generalised
ensemble 
\be 
\wp_{\al,\,\nu=0}^{(2)}(y)\ \equiv\
-\frac{\partial}{\partial y}{\cal E}_{\al,\,0}^{(2)}(y^2) \ =\
|y|\,\frac{(\al+1)}{2 \al b^2}\left(1+\frac{y^2}{4 \al b^2}
\right)^{-(\al+2)} \ . 
\label{wp1aldef} 
\ee 
These distributions
are all normalised to unity, 
\be 
\int_0^\infty dy\
\wp_{\al,\,\nu}^{(\beta)}(y)\ =\ 1\ . 
\label{wp1norm} 
\ee 
The
restriction to a Gaussian potential in the discussion above can be
lifted, as the first eigenvalue distributions for all $\nu$ are
universal, including the ratio in partition functions that we have
inserted again. Eqs. (\ref{wp1def}) and  (\ref{wp1aldef}) are
compared to the corresponding densities eqs. (\ref{bessellaw}) and
(\ref{genbessellaw}) in fig. \ref{varyal} for various values of
$\al$.

Next we give the first eigenvalue distribution for general $\nu$.
Here we directly use the most compact universal expression
\cite{DN} for $\wp_\nu^{(2)}(y)$ in WL, without making the detour
over ${\cal E}^{(2)}(y)$ \cite{DNW}, 
\be 
\wp_{\nu}^{(2)}(y) \ =\
\frac12 |y| \exp\left[-\frac14 y^2\right]\det_{1\leq i,j\leq
\nu}\left[I_{i-j+2}(|y|)\right]\ .
\label{wp1beta2nu} 
\ee 
In addition to the exponential in eq. (\ref{wp1def}) it contains a
determinant of finite size $\nu\times\nu$ over the modified
Bessel-$I$ function, which is absent at $\nu=0$. Knowing that the
properly rescaled microscopic gap probability is a function of the
form ${\cal E}^{(2)}_\nu(y;\xi)={\cal E}^{(2)}_\nu(\sqrt{\xi y/\al
b^2})$, see eq. (\ref{genbessellaw}), we obtain for the
generalised first eigenvalues distribution in terms of squared
eigenvalues 
\bea 
\wp_{\al,\,\nu}^{(2)}(y) &=&
\frac{1}{\Gamma(\al+1)} \int_0^\infty  d\xi~e^{-\xi}\
\xi^{\al}(-)\frac{\partial}{\partial y} {\cal
E}^{(2)}(y\sqrt{\xi/\al b^2})
\nn\\
&=&\frac{1}{\Gamma(\al+1)}
\int_0^\infty  d\xi~e^{-\xi}\ \xi^{\al}\frac{\xi |y|}{2\al b^2}
 \exp\left[-\frac{\xi y^2}{4\al b^2} \right]
\det_{1\leq i,j\leq
\nu}\left[I_{i-j+2}\Big(\frac{|y|}{b}\sqrt{\xi/\al}\Big) \right]\ .
\label{wp1albeta2nu} 
\eea 
For $\nu=1$ containing only one
Bessel-$I$ the integral can be performed and is given in terms of
a hypergeometric function 
\bea 
\wp_{\al,\,1}^{(2)}(y) &=&
\frac{|y|}{2\al b^2\Gamma(\al+1)} \int_0^\infty  d\xi~e^{-\xi}\
\xi^{\al+1}
 \exp\left[-\frac{\xi y^2}{4\al b^2} \right]
I_{2}\Big(\frac{y}{b}\sqrt{\xi/\al}\Big) \nn\\
&=& \frac{\Gamma(\al+3)}{|y|\Gamma(\al+1)}
\left(1+\frac{y^2}{4\al}\right)^{-(\al+3)} \
_1F_1\left(\al+3;3;\Big(1+\frac{4\al}{y^2}\Big)^{-1}\right)\ .
\label{wp1albeta2nu1} 
\eea 
It is shown in fig. \ref{varynu} (middle),
together with $\nu=2$ (right). For increasing $\nu$ however, the integral
representation eq. (\ref{wp1albeta2nu}) is more convenient.

\begin{figure}[-h]
\centerline{
\epsfig{figure=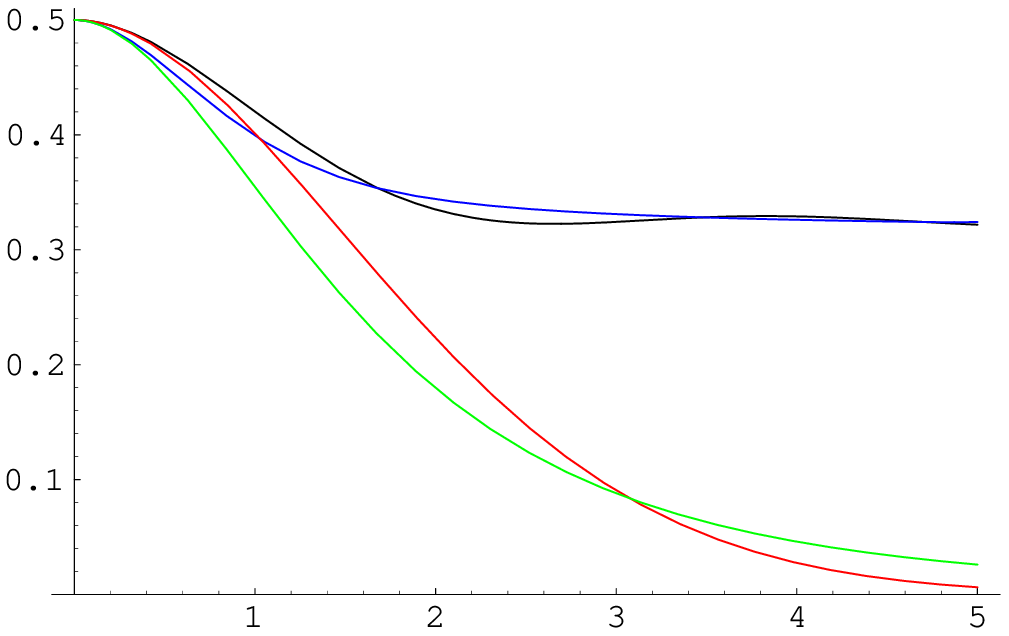
,width=13pc}
\epsfig{figure=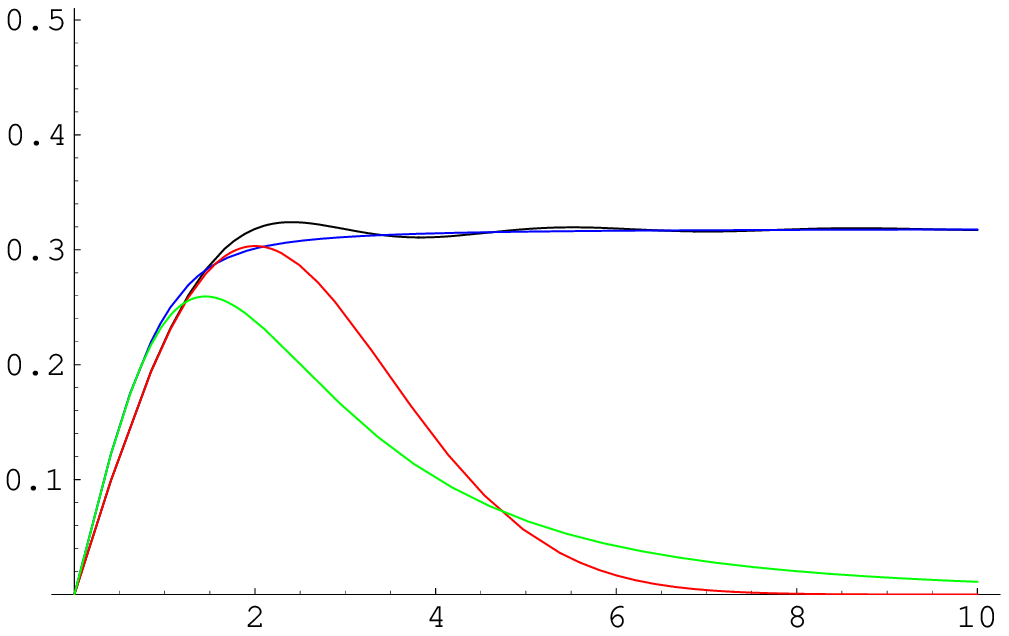
,width=13pc}
\epsfig{figure=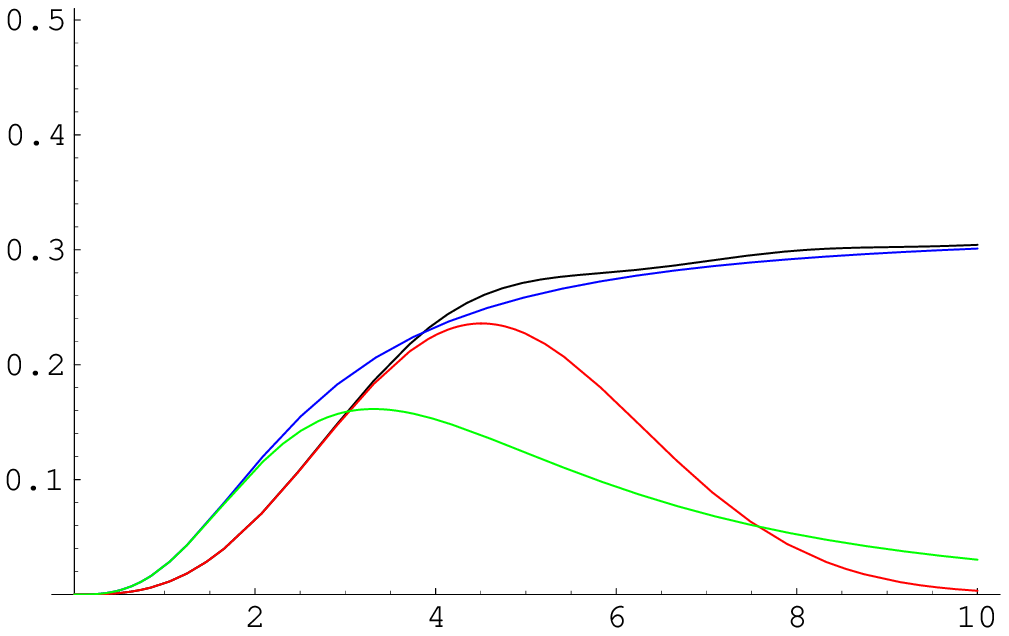
,width=13pc}
} \caption{Varying $\nu$ for $\beta=1$: the generalised
microscopic density $\vartheta_{\al,\nu}^{(1)}(y)$  eq. (\ref{rhomical1}) 
(blue) with its first eigenvalue
  (green) at
  $\al=0.1$ vs the corresponding WL Bessel density
  $\vartheta_{\nu}^{(1)}(y)$ eq. (\ref{rhomic1})
(black) and its first
  eigenvalue (red) at $\nu=0$ (left),  $\nu=1$ (middle), and $\nu=3$ (right).}
\label{varynubeta1}
\end{figure}

We now turn to $\beta=1$. Here the first eigenvalue distribution
of the WL ensembles is only known explicitly for odd values of
$\nu$ in the large-$N$ limit \cite{DN}, given by a Pfaffian with
indices running over half integers: 
\be 
\wp_{\nu}^{(1)}(y) \ = \
const.~ |y|^{(3-\nu)/2} \exp\left[-\frac18
y^2\right]\Pf_{-\frac{\nu}{2}+1\leq i,j\leq
  \frac{\nu}{2}-1}
\left[(i-j)I_{i+j+3}(|y|)\right]\ . 
\label{wp1beta1nu} 
\ee 
The
constant in front is determined by the normalisation to unity and
can be computed case by case. An exception is $\nu=0$ where the
distribution was calculated in \cite{Forrester1st} 
\be
\wp_{0}^{(1)}(y) \ =\
\frac14(2+|y|)\exp\left[-\frac{|y|}{2}-\frac{y^2}{8}\right]\ .
\label{wpbeta1nu0} 
\ee 
For $\nu=1$ and 3  we have from eq.
\eqref{wp1beta1nu} 
\bea 
\wp_{1}^{(1)}(y) &=& \frac14 |y|\exp\left[-\frac18 y^2\right]\ ,
\label{wpbeta1nu1}\\
\wp_{3}^{(1)}(y) &=& \frac12 \exp\left[-\frac18 y^2\right]I_3(|y|)\ . 
\label{wpbeta1nu3} 
\eea 
The corresponding generalised formula
for general odd $\nu$ thus reads 
\bea 
\wp_{\al,\,\nu}^{(1)}(y)
&=&const.~ \frac{1}{\Gamma(\al+1)} \int_0^\infty  d\xi~e^{-\xi}\
\xi^{\al} \sqrt{\frac{\xi}{\al b^2}} \left(\frac{|y|}{b}
\sqrt{\frac{\xi}{\al}}\right)^{(3-\nu)/2}
 \exp\left[-\frac{1}{8\al b^2} \xi y^2\right]\nn\\
&&\ \ \ \ \ \ \ \ \ \ \ \ \times\ \Pf_{-\frac{\nu}{2}+1\leq
i,j\leq \frac{\nu}{2}-1}
\left[(i-j)I_{i+j+3}\Big(\frac{|y|}{b}\sqrt{\xi/\al}\Big)\right]\ ,
\label{wp1albeta1nu} 
\eea 
up to the normalisation constant. For
the simplest examples $\nu=0,1$ and 3 displayed in  fig.
\ref{varynubeta1} we have  
\bea 
\wp_{\al,\,0}^{(1)}(y) &=&
\frac{1}{4\sqrt{\al}\,\Gamma(\al+1)b} \int_0^\infty\!
d\xi~e^{-\xi}
\xi^{\al+\frac12}\Big(2+\frac{|y|}{b}\sqrt{\frac{\xi}{\al}} \Big)
\exp\left[-\frac{|y|}{2b} \sqrt{\frac{\xi}{\al}}-\frac{\xi y^2}{8\al
b^2} \right],
\label{wp1albeta1nu0}\\
\wp_{\al,\,1}^{(1)}(y)
&=&  \frac{(\al+1)}{4 \al b^2}\,|y|
\left(1+\frac{y^2}{8 \al b^2}\right)^{-(\al+2)} \ ,
\label{wp1albeta1nu1}\\
\wp_{\al,\,3}^{(1)}(y)
&=& \frac{1}{\Gamma(\al+1)}
\int_0^\infty  d\xi~e^{-\xi}\ \xi^{\al}
 \sqrt{\frac{\xi}{\al b^2}}\frac12  \exp\left[-\frac{\xi y^2}{8\al b^2} \right]
I_{3}\Big(\frac{|y|}{b}\sqrt{\xi/\al}\Big)\ .
\label{wp1albeta1nu3}
\eea

While the density is modified only rather mildly compared to WL,
the first eigenvalue changes considerably. We have checked that
the curves converges to WL for large $\al$, where the convergence
to the density is much faster than for $\beta=2$. The case $\nu=0$
in fig. \ref{varynubeta1} left is the only example where the
microscopic densities do not vanish at $x=0$. The fact that they
both have the same limit $\frac12$ can also be seen analytically,
exploiting that $J_\nu(0)=\delta_{\nu,0}$ (see eqs.
(\ref{rhomic1}) and (\ref{rhomical1})). As mentioned above for
$\nu=2k$ with $k\in\mathbb{N}_+$ the first eigenvalue is not
available to date.

\begin{figure}[-h]
\centerline{\epsfig{figure=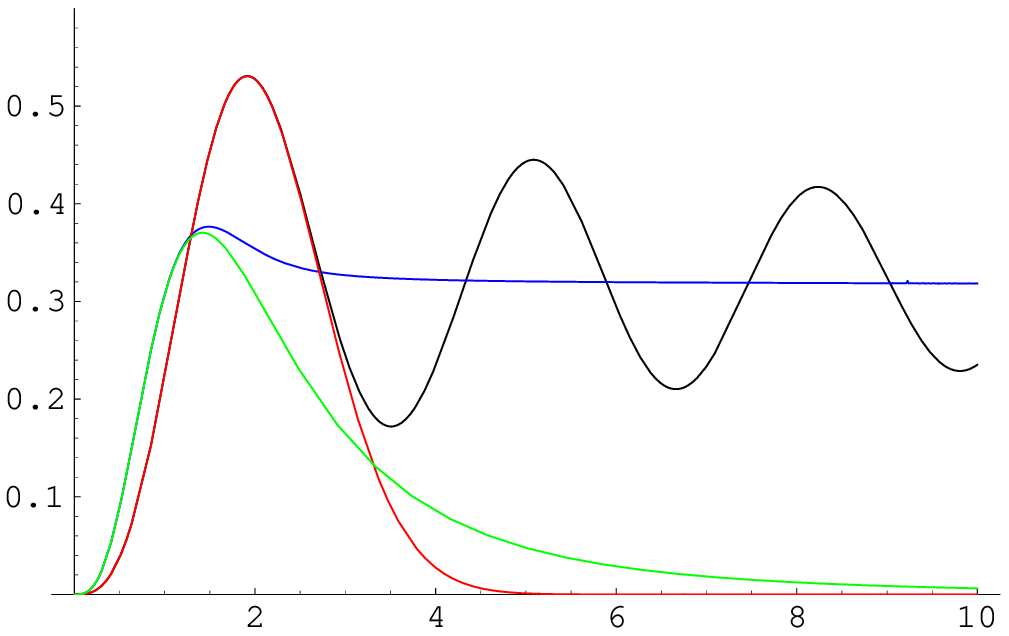
,width=13pc}
\epsfig{figure=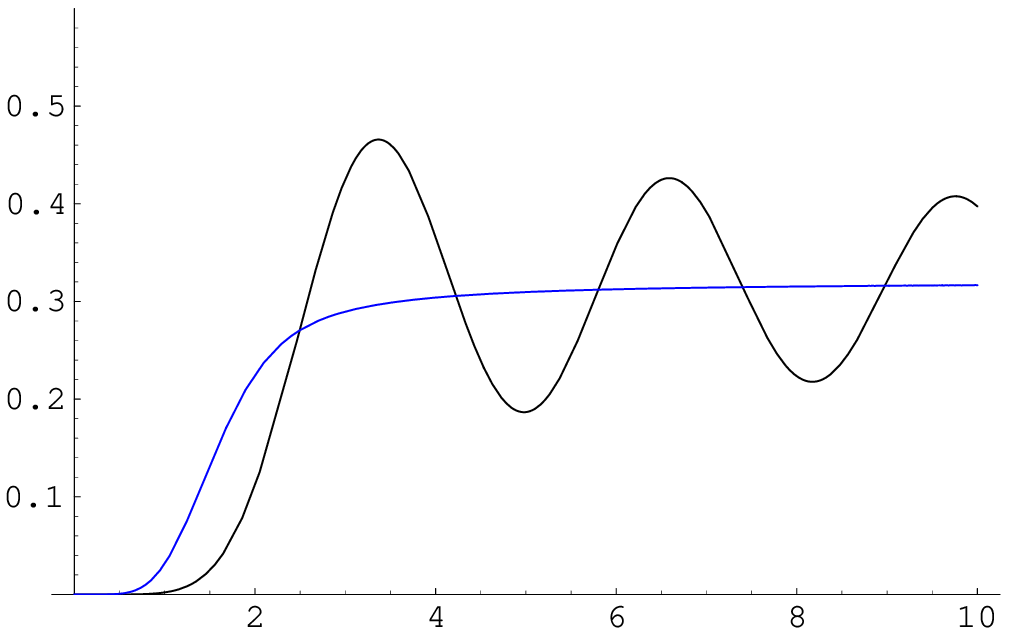
,width=13pc}
\epsfig{figure=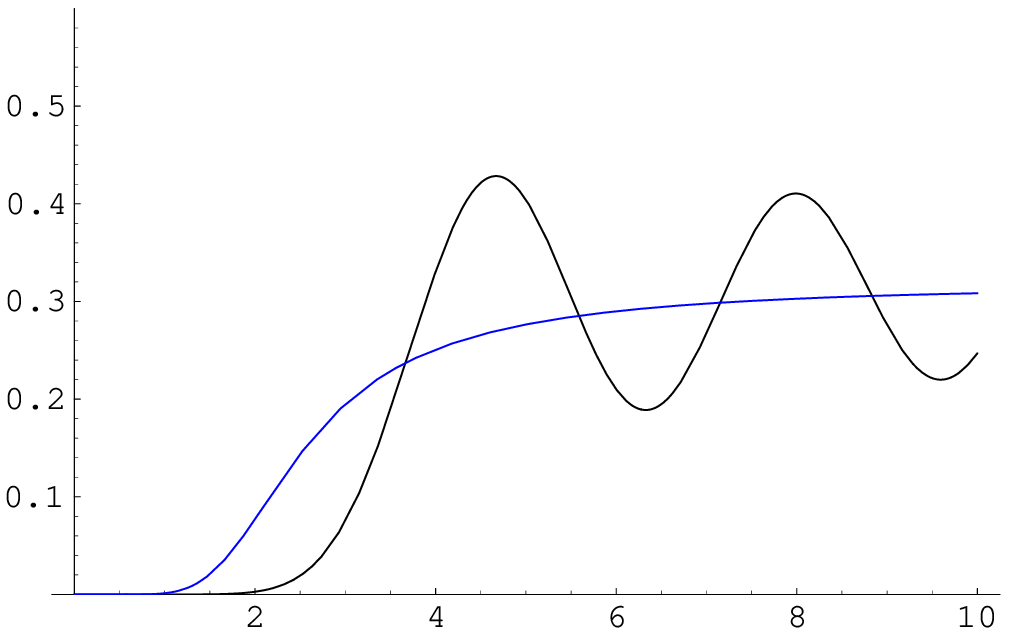
,width=13pc}
} \caption{Varying $\nu$ for $\beta=4$: the generalised
microscopic density $\vartheta_{\al,\nu}^{(4)}(y)$  eq. (\ref{rhomical4}) 
(blue) at
  $\al=0.1$ vs the corresponding WL Bessel density
  $\vartheta_{\nu}^{(4)}(y)$ eq. (\ref{rhomic4})
(black)  at $\nu=0$ (left),  $\nu=1$ (middle), and $\nu=2$
(right). For $\nu=0$ we also display the respective first
eigenvalue for the generalised (green) and WL ensemble (red).}
\label{varynubeta4}
\end{figure}

Finally we turn to $\beta=4$. In principle the result is known in
the WL ensemble, 
\be 
\wp_{0}^{(4)}(y) \ = \ const.~\ |y|^{\nu
+\frac32} \exp\left[-\frac12 y^2\right] Z_{3/2}(\{|y|\}_{2\nu+1})\ .
\ee 
Here $Z_{3/2}(\{y\}_{2\nu+1})$ is the large-$N$ matrix model partition
function at topological charge $3/2$ with $2\nu+1$ degenerate
masses at value $y$,  and we refer to \cite{DN} for a 
more detailed discussion of these objects.
This partition function is generally
known explicitly only for an even number of masses, 
except at $\nu=0$. There we have
\be 
\wp_{0}^{(4)}(y) \ =\ \frac12\sqrt{2\pi}\ |y|^{\frac32}
\exp\left[-\frac12 y^2\right] I_{3/2}(y) \ =\ |y|\
\Big(\cosh(y)-\frac1y\sinh(y)\Big) \exp\left[-\frac12 y^2\right]\ . 
\label{wp1beta4nu0} 
\ee 
Thus the generalised distribution
depicted in fig. \ref{varynubeta4} is given by 
\be
\wp_{0}^{(4)}(y) \ =\  \frac{1}{\al\Gamma(\al+1)b^2} \int_0^\infty
d\xi~e^{-\xi}\ \xi^{\al+1} \left(
|y|\cosh\Big(\frac{y}{b}\sqrt{\xi/\al}\Big)
-b\sqrt{\frac{\al}{\xi}}\sinh\Big(\frac{|y|}{b}\sqrt{\xi/\al}\Big)
\right)\exp\left[-\frac{\xi y^2}{2\al b^2} \right]\ .
\label{wp1albeta4nu0} 
\ee

We note that the oscillations get smoothed out as already observed at
$\beta=2$. The convergence in $\al$ to the WL quantity is again slow, as
for $\beta=2$.

%%%%%%%%%%%%%%%%%%%%%%%%%%%%%%%%%%%%%%%%%%%%%%%%%%%%%%%%%%%%%%%%%%%%%%%%%%

\sect{The Wigner surmise in the bulk}
\label{spacing}

In this section we study the spacing distribution between
eigenvalues in the bulk in our generalised model. In contrast to
the previous sections we do not take $N$ to be large, but use the
$N=2$ results, following the original idea of Wigner.

In the Wigner-Dyson ensembles with a Gaussian potential, the
spacing distribution has the simple form 
\be 
{\cal P}_{W\!D}^{(\beta)}(s)\ = \ A s^\beta e^{-B s^2}\ , 
\ee 
known as Wigner's surmise. The known constants $A,B$ follow from
normalisation and can be found e.g. in \cite{GMW}. For the WL case
the corresponding expression can be computed from the jpdf, where we introduce
$\bar{\nu}\equiv\frac{\beta}{2}(\nu+1)-1$:
\begin{equation}\label{GapDistribution}
{\cal  P}^{(\beta)}(s)=C
  s^{\beta+\bar{\nu}+1/2}K_{1/2+\bar{\nu}}(n\beta s)
\end{equation}
where $K_\mu(x)$ is a modified Bessel function and the constant $C$
is given by:
\begin{equation}\label{CWL}
  C=\left(2^{-1/2+\beta+\bar{\nu}}(n\beta)^{-3/2-\beta-\bar{\nu}}
\Gamma\left(\frac{1+\beta}{2}\right)
\Gamma\left(1+\bar{\nu}+\frac{\beta}{2}\right)\right)^{-1}
\end{equation}

Only for $\bar{\nu}=0$, one recovers the WL Wigner's surmise, using
$K_{1/2}(x)=\sqrt{\pi/2x}\,e^{-x}$ (apart from squared variables in
the exponent there). For general $\bar{\nu}$ one easily gets an
exponential decay for large spacing $\sim s^{\beta+\bar{\nu}}e^{-n\beta
s}$, from the asymptotic expression of the Bessel function, 
$\lim_{x\to\infty}K_{\mu}(x)\sqrt{\pi/2x}\,e^{-x}$.
However, we expect that only the $\bar{\nu}=0$ expression will lead to a
good approximation of the infinite-$N$ case.

Conversely, for the generalised model the spacing distribution has
a different expression:
\begin{equation}\label{GapDistributionGen}
{\cal P}_\gamma^{(\beta)}(s)=C_\gamma
s^{\beta+\bar{\nu}}\left(\frac{\gamma}{2n\beta}
+\frac{s}{2}\right)^{\bar{\nu}+1-\gamma} 
~_2
F_1\left(-\bar{\nu},\bar{\nu}+1;-\bar{\nu}+\gamma;\frac{1}{2}-\frac{\gamma}{2n
  \beta s}\right)
\end{equation}
where $_2 F_1(a,b;c;z)$ is a hypergeometric function. The
constant $C_\gamma$ can be computed as
\begin{equation}\label{Cgamma}
  C_\gamma=\frac{2^{1-\gamma}B(\bar{\nu}+1,-2\bar{\nu}-1+\gamma)\Gamma(\gamma)
\Gamma(1+\beta/2)}{\Gamma(\bar{\nu}+1)\Gamma(1+\beta)\Gamma
(\bar{\nu}+1+\beta/2)
    \Gamma(\gamma-2-\beta-2\bar{\nu})}
\left(\frac{n\beta}{\gamma}\right)^{2+2\bar{\nu}+\beta-\gamma}
\end{equation}
and $B(\alpha,\beta)$ is a Beta function.
The large-$s$ decay is now
a pure power law $\sim s^{-(\varpi+1)}$ where we have defined
\be
\varpi\ \equiv\ \gamma-\beta-2\bar{\nu}-2\ > \ 0\ ,
\label{converge2}
\ee
required to be positive for
convergence\footnote{This condition equals
  eq. (\ref{converge}) derived for the partition function at $N=2$.}.

For $\bar{\nu}=0$, the spacing distribution takes a much simpler form:
\begin{equation}\label{GapDistributionGenNu0}
{\cal P}_\gamma^{(\beta)}(s)\
=\ \frac{n(n\beta/\gamma)^{\beta}\Gamma(\gamma-1)}{\gamma
\Gamma(\beta)\Gamma(\gamma-\beta-2)}\
s^\beta\left(1+\frac{sn\beta}{\gamma}\right)^{1-\gamma}\ .
\end{equation}
It agrees with the corresponding quantity in the generalised WD ensembles
found in \cite{Adel}. 
In this particular case, we can compute the mean level spacing
explicitly:
\begin{equation}\label{MeanLevelSpacing}
  \svev_\gamma=\int_0^\infty s\
  {\cal P}_\gamma^{(\beta)}(s)ds=\left(\frac{\gamma}{n\beta}\right)
\frac{1+\beta}{\gamma-\beta-3}\ ,
\end{equation}
which converges to the WL Wigner surmise value for $\gamma\to\infty$:
\begin{equation}\label{MeanLevelSpacingWigner}
  \svev=\int_0^\infty s\
  {\cal P}^{(\beta)}(s)ds =\frac{1+\beta}{n\beta}\ .
\end{equation}

\begin{figure*}[htb]
\begin{center}
  \unitlength1.0cm
\epsfig{file=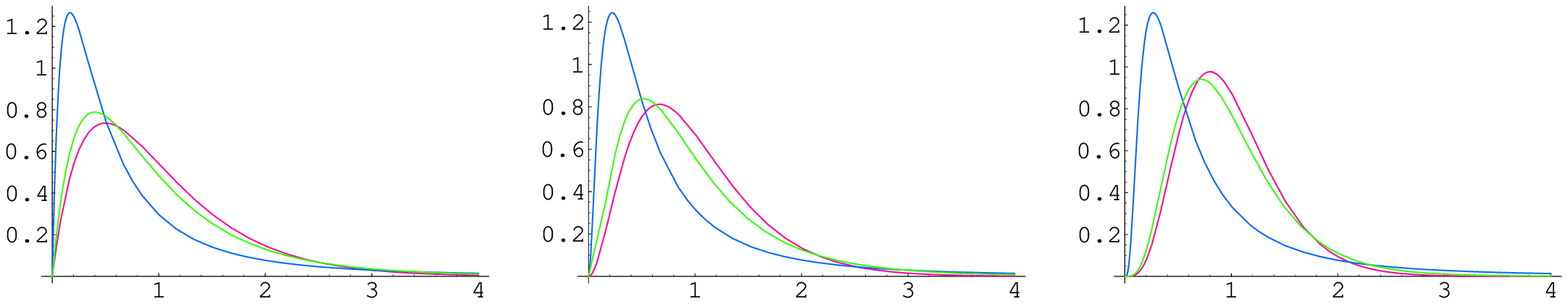,width=18cm}
  \caption{
    \label{WS}
    Comparison between $\hat{{\cal P}}_\gamma^{(\beta)}(x)$ (blue, green) and
    $\hat{{\cal P}}^{(\beta)}(x)$ (red), for $\beta=1,2,4$
    (from left to right). The $\gamma$ value for the blue curves is chosen is
    chosen in such a way that the combination $\varpi$ is kept
    constant to the value $2$. The green curves have value
    $\ga=12,12,25$ from left to right, and correctly approach the limiting WL
    curve. }
\end{center}
\end{figure*}

After defining the rescaled quantities having mean spacing $1$,
\begin{align}\label{Ptildes}
\hat{{\cal P}}_\gamma^{(\beta)}(x)
&=\svev_\gamma {\cal P}_\gamma^{(\beta)}\Big(\svev_\gamma x\Big)\\
\hat{{\cal P}}^{(\beta)}(x) &=\svev {\cal P}^{(\beta)}
\Big(\svev x\Big)
\end{align}
we can compare the curves for all three
$\beta$ at $\bar{\nu}=0$ (and $n=1$), in fig. \ref{WS}. The power-law tail
modification compared to the standard WL spacing distribution
is evident in the plots. Because in the large-$N$ limit $\ga$ scales with $N$,
we keep the combination $\varpi$ in eq. \eqref{converge2} fixed to be able to
compare to spacing distributions at large-$N$.

Both in this and the previous section the power-law tail of the macroscopic
density is seen to persist on the microscopic level of the mean level-spacing.
It would be very interesting to confirm this on real data sets.

%%%%%%%%%%%%%%%%%%%%%%%%%%%%%%%%%%%%%%%%%%%%%%%%%%%%%%%%%%%%%%%%%%%%%%%%%%%
\sect{Conclusions and outlook}\label{con}

\begin{figure}[htb]
\begin{center}
\includegraphics[bb = 53 189 546 589,totalheight=0.35\textheight]{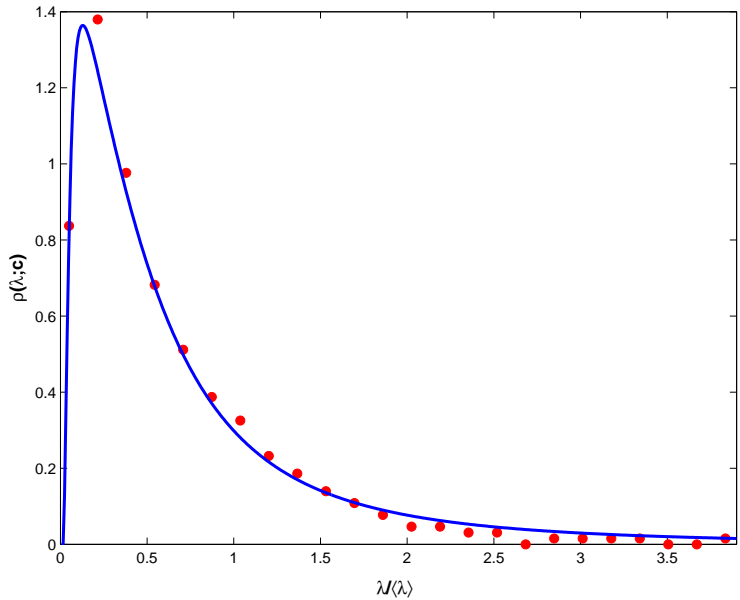}
\caption{Comparison between the rescaled eigenvalue distribution from
  financial data \cite{AFV}
and the macroscopic density $\rho_\al(x)$ for the generalised model eq. 
  \eqref{rhoalc<1}, in red dots and solid blue respectively.
The best fit gives a value of $\al\approx 0.95$, which corresponds
to a power-law decay as $\rho_\al(x)\sim x^{-2.95}$.} \label{Comparison}
\end{center}
\end{figure}

We have introduced a generalisation of all three ensembles of random matrices
called Wishart-Laguerre or chiral ensembles, replacing
the exponential of a non-Gaussian potential by a fat-tailed
distribution with parameter $\ga$.
In the limit $\ga\to\infty$ we
can recover the exponential weight and thus the standard ensembles. This
modification lead to the appearance of correlations with a power-law, 
governed by a single parameter. Such a behaviour is found in many
systems in nature, e.g. in the wide area of complex networks. WL
ensembles are often used in comparing to eigenvalues from
covariance matrices of real data sets. To illustrate the potential
of our generalisation we show a comparison to financial data in fig.
\ref{Comparison}. The eigenvalues of the covariance matrix from
time series of stock data clearly show a power-law behaviour. These
are well described by the generalised Mar\v{c}enko-Pastur density,
refining previous comparisons to the standard Mar\v{c}enko-Pastur
law.

The solution of our generalised models relied heavily on the
possibility of writing them as an integral transform of the standard
WL ensembles. This generalisation is thus in the spirit of
super-statistics where other models have been constructed already.
The virtue of our model is its invariance for all three symmetry
classes, allowing to go to an eigenvalues basis and to study
universality. We could show that the generalised macroscopic density 
which was known in the Gaussian case is
only weakly universal.

In contrast all microscopic densities are universal
under any invariant deformations by polynomial potentials. This
macroscopic/microscopic dichotomy should not come as a surprise,
being observed previously for the restricted trace ensembles.

We exploited the linear relation to standard WL to solve our 
model exactly at finite-$N$ for any polynomial potential, using
the formalism of orthogonal polynomials therein. In the subsequent
large-$N$ double scaling limit, where $\ga$ is scaled with $N$, we
derived all density correlations in the macroscopic limit for quadratic and
rectangular matrices, and in the
microscopic limit at the hard edge for all three values of 
$\beta$.  Here we have
mainly focused on the spectral density itself and the first
eigenvalue distribution. The general formalism for higher density
or higher individual eigenvalue correlation functions was provided and is
straightforward to use if such quantities will be needed.

While the hard edge was solved exhaustively, persisting in our
model for asymptotically quadratic matrices, we only provided a
Wigner surmise in the bulk. Here, more detailed properties of correlation
functions could be investigated, including a possible
generalisation of the Tracy-Widom distribution at the soft edges.
This is left for future investigations.

\indent

\noindent \underline{Acknowledgements}: We wish to thank Oriol
Bohigas, Mauricio Pato and Zdzis\l aw Burda for fruitful
discussions on their related models, Giulio Biroli, Matteo Marsili and Luca
Dall'Asta for comments, and Poul Damgaard for useful
correspondence. We are also indebted to Jonit Fischmann for
sharing her data and a related collaboration, as well as to Leonid
Shifrin for help with Mathematica. Financial support by EPSRC
grant EP/D031613/1, European Network ENRAGE MRTN-CT-2004-005616
(G.A.), and European Union Marie Curie Programme NET-ACE (P.V.) is
gratefully acknowledged.

%%%%%%%%%%%%%%%%%%%%%%%%%%%%%%%%%%%%%%%%%%%%%%%%%%%%%%%%%%%%%%%%%%%%%%%%
\begin{appendix}

\sect{Partition function and first moment of the Gaussian models}
\label{sphere}

The purpose of this appendix is threefold. First, we derive the
condition under which our generalised model defined in eq.
\eqref{Pga} is convergent. Second, we compute both the generalised
and standard Gaussian partition functions in order to determine
their $\xi$-dependent ratio needed in the computation of all
eigenvalue correlation functions. Third, we compute the first
moment as a function of $N$, $\nu$, $\beta$ and $\ga$ which is
needed for the rescaling of the eigenvalues in the large-$N$
limit.

All three steps will be performed by changing variables from independent
matrix elements or its eigenvalues
to radial coordinates, following \cite{ACMV}.

The generalised partition function reads in terms of eigenvalues
\bea 
\mathcal{Z}_{\ga} &=&  \int_0^\infty \prod_{i=1}^N d\la_i\
\prod_{i=1}^N\lambda_i^{\frac12\beta(\nu+1)-1}
\prod_{j>k}^N|\lambda_j-\lambda_k|^\beta
\frac{1}{\left(1+\frac{n\beta}{\gamma}\sum_{i=1}^NV(\lambda_i)\right)^{\gamma}}
\\
&=& \int_{\Omega(N)}da_N \int_0^\infty dr r^{N-1}
r^{N(\frac12\beta(\nu+1)-1)}
\prod_{i=1}^N\left(\frac{\lambda_i}{r}\right)^{\frac12\beta(\nu+1)-1}
r^{\frac{N(N-1)}{2}\beta}
\frac{\prod_{j>k}^N\left|\frac{\lambda_j}{r}-\frac{\lambda_k}{r}\right|^\beta}
{\left(1+\frac{n\beta}{\gamma}\sum_{i=1}^NV(r\frac{\lambda_i}{r})
\right)^{\gamma}}\ . \nn 
\eea 
Here we have changed to radial
coordinates of the $N$-component vector of the eigenvalues
$(\la_1,\ldots,\la_N)$, and $da_N$ denotes the angular integration
over the $N$-dimensional unit sphere $\Omega(N)$. The eigenvectors
$e_i=\la_i/r$ of norm unity span $\Omega(N)$ and no longer depend
on the radius. Collecting all powers of $r$ in the numerator and
comparing to the leading power of the denominator ${d\ga}$ at
large $r\gg1$, the integral only converges if the following
inequality holds: 
\be 
\frac{\beta}{2}N(N+\nu)-1-\ga d\ <\ -1 \ ,
\ee 
which is exactly eq. \eqref{converge}.

Next we compute the partition function $\mathcal{Z}(\xi)$, where for the rest
of this appendix we restrict ourselves to the Gaussian
potential $V(\la)=\la$. The same steps can be taken
for a purely monic potential
$V(\la)=\la^d$ as well.

In principle we could repeat the same calculation in terms of
eigenvalues as above, but it will be more instructive to
start directly from the matrix elements: 
\bea 
\mathcal{Z}(\xi) &=&
\int d\mathbf{X} \exp\left[-\xi \frac{n\beta}{\ga}
\Tr\mathbf{X}^\dag\mathbf{X}\right]
\nn\\
&=& \int_{\Omega(\beta N(N+\nu))}da_{\beta N(N+\nu)}
\int_0^\infty drr^{\beta N(N+\nu)-1}
\exp\left[-\xi \frac{n\beta}{\ga}\,r^2\right]
\nn\\
&=& \frac12 \left(\frac{\ga}{\xi
n\beta}\right)^{\frac{\beta}{2}N(N+\nu)} \Gamma\left(
\frac{\beta}{2}N(N+\nu)\right) \int_{\Omega(\beta
N(N+\nu))}da_{\beta N(N+\nu)}\ . \label{Zxisphere}
\eea
Here we
have used radial coordinates for the $\beta N(N+\nu)$ component
vector of all independent matrix elements $\mathbf{X}_{ij}$, with squared norm
$r^2=\Tr\mathbf{X}^\dag\mathbf{X}$. We don't need to compute the
angular integral explicitly as it cancels out below.

If we insert the result eq. \eqref{Zxisphere} into the relation
\eqref{Zga} we immediately obtain
\bea
\mathcal{Z}_{\ga} &=&
\frac{1}{\Gamma(\gamma)}\int_0^\infty
  d\xi~e^{-\xi}\ \xi^{\gamma-1} \mathcal{Z}(\xi)
\label{Zgasphere}\\
&=& \frac{1}{2\Gamma(\ga)} \left(\frac{\ga}{\beta
n}\right)^{\frac{\beta}{2}N(N+\nu)} \Gamma\left(\ga-
\frac{\beta}{2}N(N+\nu)\right) \Gamma\left(
\frac{\beta}{2}N(N+\nu)\right) \int_{\Omega(\beta
N(N+\nu))}da_{\beta N(N+\nu)}\ . \nn 
\eea 
Combining the last two
equations we arrive at 
\be
\frac{\mathcal{Z}(\xi)}{\Gamma(\ga)\mathcal{Z}_{\ga}}\ =\
\frac{\xi^{-\frac{\beta}{2}N(N+\nu)}}{ \Gamma\left(\ga-
\frac{\beta}{2}N(N+\nu)\right)}\ , 
\label{ZratioA} 
\ee 
the ratio
of the two Gaussian partition functions at finite values of $N$,
$\nu$ and $\ga$ valid for all three $\beta$.

In the last step of this appendix we compute the mean eigenvalue
position in the Gaussian model. It can be either defined through
the spectral density, see eq. \eqref{vevla}, or in terms of the
first moment, where we start with the standard WL ensembles:
\bea
\langle\la(\xi)\rangle &=& \frac{1}{\mathcal{Z}(\xi)}
 \int d\mathbf{X} \frac1N\Tr\left(\mathbf{X}^\dag\mathbf{X}\right)
\exp\left[-\xi \frac{n\beta}{\ga}
\Tr\mathbf{X}^\dag\mathbf{X}\right]
\nn\\
&=&\frac{1}{N\mathcal{Z}(\xi)}\int_{\Omega(\beta N(N+\nu))}da_{\beta N(N+\nu)}
\int_0^\infty drr^{\beta N(N+\nu)-1}r^2
\exp\left[-\xi \frac{n\beta}{\ga}\,r^2\right]\nn\\
&=& \frac{\ga}{2n\xi}\,(N+\nu)\ . 
\label{vevWL} 
\eea 
Note that the
$\beta$-dependence has cancelled out. We can immediately use this
result to compute the same quantity for the generalised Gaussian
model, 
\bea 
\langle\la\rangle_\ga &=& \frac{1}{\mathcal{Z}_\ga}
 \int d\mathbf{X} \frac{\frac1N\Tr\left(\mathbf{X}^\dag\mathbf{X}\right)}
{\left( 1+\frac{n\beta}{\ga}\Tr\mathbf{X}^\dag\mathbf{X}\right)^\ga}
\nn\\
&=& \frac{1}{N\mathcal{Z}_\ga\Gamma(\ga)}
\int_0^\infty
  d\xi~e^{-\xi}\ \xi^{\gamma-1} \mathcal{Z}(\xi)\langle\la(\xi)\rangle\nn\\
&=& \frac{\ga(N+\nu)}{2n\left(\ga-\frac{\beta}{2}N(N+\nu)-1
\right)}. 
\label{vevgen}
\eea
This result is used in the rescaling
of both large-$N$ limits. As a check it reduces to eq.
\eqref{vevWL} in the limit $\ga\to\infty$, with a weight
$\exp[-n\beta r^2]$.

%%%%%%%%%%%%%%%%%%%%%%%%%%%%%%%%%%%%%%%%%%%%%%%%%%%%%%%%%%%%%%%%%%%
\sect{Explicit $\beta=2$-solution for all $k$-point densities at
finite $N$ and $\gamma$}\label{beta2N}

In this appendix we present all details for the solution of the
generalized WL ensemble with unitary invariance
$\beta=2$ and Gaussian potential $V(\la)=\la$. In this case the
orthogonal polynomials of the WL ensemble are know to be Laguerre,
allowing for an explicit solution at finite $N$ and finite $\ga$.

The orthogonal polynomials in eq. \eqref{OPdef} read for the
weight function $\exp[-\xi\frac{2n}{\ga}\la]$
\be
P_k(\la)\ =\ (-)^kk!\, \left(\frac{\ga}{2n\xi}\right)^k L_k^\nu
\left(\frac{2n\xi}{\ga}\la\right)\ , 
\ee 
with norms 
\be 
h_k\ =\
k!\,(k+\nu)!\,\left(\frac{\ga}{2n\xi}\right)^{2k+\nu+1}\ .
\label{norms}
\ee
Here the Laguerre polynomials are defined as
usual
\begin{equation}\label{DefinitionLaguerre}
L_k^\nu(z)\ =\ \sum_{j=0}^k (-1)^j \binom{k+\nu}{k-j}\frac{z^j}{j!}\ \ ,\ \
\mbox{with}\ \  L_k^\nu(z)^\prime\ =\ -L_{k-1}^{\nu+1}(z)\ .
\end{equation}
Using eq. \eqref{ZWLresult} we can immediately read off the WL partition
function from the norms,
\be
\mathcal{Z}(\xi)\ =\
\left(\frac{\ga}{2n\xi}\right)^{N\nu+N^2}
\prod_{k=0}^{N-1}(k+1)!\,(k+\nu)!\ .
\ee
As the next step we can
compute the partition function eq. \eqref{Zga} given by
\bea
\mathcal{Z}_\ga &=& \frac{1}{\Gamma(\gamma)}\int_0^\infty
  d\xi~e^{-\xi}\ \xi^{\gamma-1}\
 \left(\frac{\ga}{2n\xi}\right)^{N\nu+N^2}
\prod_{k=0}^{N-1}(k+1)!\,(k+\nu)!
\nn\\
&=& \left(\frac{\ga}{2n}\right)^{N\nu+N^2}
\frac{\Gamma(\ga-N(N+\nu))}{\Gamma(\ga)}
\prod_{k=0}^{N-1}(k+1)!\,(k+\nu)! \ . \label{ZgammaNuneq0}
\eea
This leads to the following ratio needed for example inside the relation
\eqref{Rkgarel}
\be
\frac{\mathcal{Z}(\xi)}{\Gamma(\gamma)\mathcal{Z}_\ga} \ =\
\xi^{-N(N+\nu)}\frac{1}{\Gamma(\ga-N(N+\nu))}\ .
\ee
It confirms
independently part of the result from the previous appendix, eq.
\eqref{ZratioA} for $\beta=2$.

The spectral density for finite $N$ follows by inserting this
ratio as well as the standard Laguerre density at finite-$N$,
\be
R(\lambda;\xi)\ =\ \la^\nu\ e^{-\xi\frac{2n}{\ga}\la}
\sum_{k=0}^{N-1}\frac{k!}{(k+\nu)!}\left(\frac{2n\xi}{\gamma}\right)^{\nu+1}
L_k^{\nu}\left(\frac{2n\xi}{\gamma}\la\right)^2\ ,
\label{RGN}
\ee
into \eqref{Rkgarel}:
\be
R_\ga(\lambda)\ =\
\frac{1}{\Gamma(\ga-N(N+\nu))} \int_0^\infty d\xi\ e^{-\xi}\
\xi^{\gamma-1-N(N+\nu)}R(\lambda;\xi)\ .
\label{Rgainsert}
\ee
With the help of eq. \eqref{DefinitionLaguerre} we can derive and
simplify the Christoffel-Darboux identity for Laguerre polynomials
of equal arguments
\begin{equation}
\label{ChristDarb}
\sum_{k=0}^{N-1}\frac{k!}{\Gamma(k+1+\nu)}L_k^{\nu}(y)^2=
\frac{N!}{\Gamma(N+\nu)}\left[L_{N-1}^{\nu}(y)
L_{N-1}^{\nu+1}(y)-L_{N}^{\nu}(y)L_{N-2}^{\nu+1}(y)\right]\ .
\end{equation}
We thus arrive at our final result for the generalised density at
finite $N$ and $\ga$: 
\bea 
R_\ga(\lambda)
&=&\frac{N!}{\Gamma(\gamma-N^2-N\nu)\Gamma(N+\nu)}
\left(\frac{2n}{\gamma}\right)^{\nu+1}\lambda^\nu \int_0^\infty
d\xi~e^{-\xi\left(1+\frac{2n}{\gamma}\lambda\right)}
\xi^{\gamma-N^2-N\nu+\nu}\nn\\
&&\times \left[L_{N-1}^{\nu}\left(\frac{2n\xi}{\gamma}\lambda\right)
L_{N-1}^{\nu+1}\left(\frac{2n\xi}{\gamma}\lambda\right)
 -L_{N}^{\nu}\left(\frac{2n\xi}{\gamma}\lambda\right)
L_{N-2}^{\nu+1}\left(\frac{2n\xi}{\gamma}\lambda\right)\right]\ .
\label{RgaGN}
\eea
This single integral over an exponential times polynomials can
be performed explicitly, at the expense of a double sum. Since this
equivalent result is not very illuminating or useful for the asymptotic
we do not display it here.

Proceeding along the same lines as above we can write down the general
result for the $k$-point density correlations functions as they
follow from eq. \eqref{Rkgaresult}
\bea
R_\ga(\lambda_1,\ldots,\la_k)
&=&\frac{N!^k\left(\frac{n}{\gamma}\right)^{k\nu}}
{\Gamma(\gamma-N^2-N\nu)\Gamma(N+\nu)^k}
\prod_{j=1}^k\lambda_j^\nu \int_0^\infty d\xi~
e^{-\xi\left(1+\frac{2n}{\gamma}\sum_{j=1}^k\lambda_j\right)}
\xi^{\gamma-1-N^2-(N-k)\nu}\nn\\
&&\times
\det_{1\leq i,j\leq k}
\left[\frac{
\left(
L_{N}^{\nu}\left(\frac{2n\xi}{\gamma}\lambda_i\right)
L_{N-1}^{\nu}\left(\frac{2n\xi}{\gamma}\lambda_j\right)
 -L_{N}^{\nu}\left(\frac{2n\xi}{\gamma}\lambda_j\right)
L_{N-1}^{\nu}\left(\frac{2n\xi}{\gamma}\lambda_i\right)\right)
}{\la_j-\la_i}\right]\ .
\nn\\
\label{RkgaGauss}
\eea

\begin{figure*}[htb]
\begin{center}
\centerline{
\epsfig{file=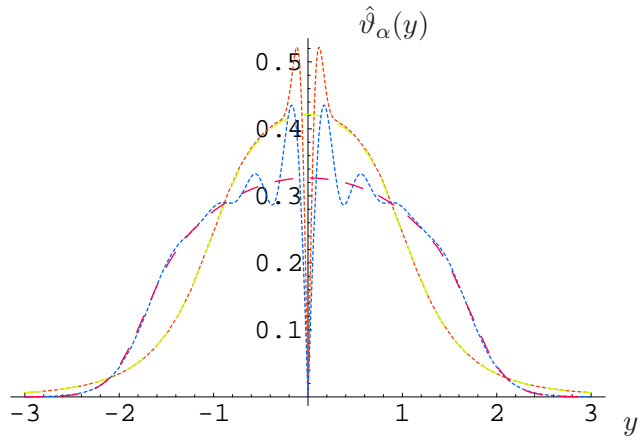,width=20pc}
\put(-100,150){$\hat{\vartheta}_\al(y)$}
\put(0,0){$y$}
}  
\caption{
    \label{mac2N}
The macroscopic generalised semi-circle density 
$\vartheta_\al(x)$ eq. \eqref{Rmap}
for $\al=1.02$ and $14$ (green and dashed red),
compared with the finite $N=4$ result $\hat{\vartheta}_\al(x)$
(blue and dash-dotted orange). }
\end{center}
\end{figure*}

In order to compare the finite-$N$ result \eqref{RgaGN} with the
macroscopic $\ga$- and $N$-independent density $\vartheta_\al(x)$ eq. 
(\ref{Rmap}), we adopt the following procedure:
\begin{enumerate}
  \item We rescale $R_\ga(\lambda)$ with mean value
  $\langle\lambda\rangle_\ga$ eq. (\ref{vevla}) and normalise to $1$:
  $\hat{\rho}_\ga(x)\equiv N^{-1}\langle\lambda\rangle_\ga
  R_\ga\left(\langle\lambda\rangle_\ga x\right)$.
  \item Next, we express $\ga$ as a function of $\al$ and $N$,
  $\ga=\al+N(N+\nu)+1$, and pass to squared
  variables:
  $\hat{\vartheta}_\al(x)\equiv|x|\hat{\rho}_{\ga}(x^2)$.
  \item Then, we compare $\hat{\vartheta}_\al(x)$ and
${\vartheta}_\al(x)$ for
  $\nu=0$ in Fig. \ref{mac2N}. The agreement is already very good 
  for $N=4$, apart from the region close to the origin.
\end{enumerate}

\end{appendix}

\end{document}